\documentclass[aps,pra,twocolumn,nopacs,superscriptaddress,groupedaddress]{revtex4}  
\usepackage{graphicx}  
\usepackage{dcolumn}   
\usepackage{bm}        
\usepackage{amssymb}   
\usepackage{times}
\usepackage{epstopdf}

\usepackage{color}
\usepackage{hyperref} 
\makeatletter

\@ifundefined{textcolor}{}
{%
 \definecolor{BLACK}{gray}{0}
 \definecolor{WHITE}{gray}{1}
 \definecolor{RED}{rgb}{1,0,0}
 \definecolor{GREEN}{rgb}{0,1,0}
 \definecolor{BLUE}{rgb}{0,0,1}
 \definecolor{CYAN}{cmyk}{1,0,0,0}
 \definecolor{MAGENTA}{cmyk}{0,1,0,0}
 \definecolor{YELLOW}{cmyk}{0,0,1,0}
}

\makeatother

\begin{document}
\let\vaccent=\v{
\global\long\def\gv#1{\ensuremath{\mbox{\boldmath\ensuremath{#1}}}}
\global\long\def\uv#1{\ensuremath{\mathbf{\hat{#1}}}}
\global\long\def\abs#1{\left| #1 \right|}
\global\long\def\avg#1{\left< #1 \right>}
\let\underdot=\d{
\global\long\def\dd#1#2{\frac{d^{2}#1}{d#2^{2}}}
\global\long\def\pd#1#2{\frac{\partial#1}{\partial#2}}
\global\long\def\pdd#1#2{\frac{\partial^{2}#1}{\partial#2^{2}}}
\global\long\def\pdc#1#2#3{\left( \frac{\partial#1}{\partial#2}\right)_{#3}}
\global\long\def\op#1{\hat{\mathrm{#1}}}
\global\long\def\ket#1{\left| #1 \right>}
\global\long\def\bra#1{\left< #1 \right|}
\global\long\def\braket#1#2{\left< #1 \vphantom{#2}\right| \left. #2 \vphantom{#1}\right>}
\global\long\def\matrixel#1#2#3{\left< #1 \vphantom{#2#3}\right| #2 \left| #3 \vphantom{#1#2}\right>}
\global\long\def\av#1{\left\langle #1 \right\rangle }
 \global\long\def\com#1#2{\left[#1,#2\right]}
\global\long\def\acom#1#2{\left\{  #1,#2\right\}  }
\global\long\def\grad#1{\gv{\nabla} #1}
\let\divsymb=\div 
\global\long\def\div#1{\gv{\nabla} \cdot#1}
\global\long\def\curl#1{\gv{\nabla} \times#1}
\let\baraccent=\={

\title{Stabilizing strongly correlated photon fluids with non-Markovian reservoirs}
\author{Jos\'e Lebreuilly}
\email{jose.lebreuilly@unitn.it}
\affiliation{INO-CNR BEC Center and Dipartimento di Fisica, Universit\`a di Trento, I-38123 Povo, Italy}
\author{Alberto Biella} 
\affiliation{Universit\'{e} Paris Diderot, Sorbonne Paris Cit\'{e}, Laboratoire Mat\'{e}riaux et Ph\'{e}nom\`{e}nes Quantiques, CNRS-UMR 7162, 75013 Paris, France} 
\affiliation{NEST, Scuola Normale Superiore \& Istituto Nanoscienze-CNR, I-56126 Pisa, Italy}
\author{Florent Storme}
\affiliation{Universit\'{e} Paris Diderot, Sorbonne Paris Cit\'{e}, Laboratoire Mat\'{e}riaux et Ph\'{e}nom\`{e}nes Quantiques, CNRS-UMR 7162, 75013 Paris, France}
\author{\\Davide Rossini} 
\affiliation{Dipartimento di Fisica, Universit\`{a} di Pisa and INFN, Largo Pontecorvo 3, I-56127 Pisa, Italy}
\affiliation{NEST, Scuola Normale Superiore \& Istituto Nanoscienze-CNR, I-56126 Pisa, Italy}
\author{Rosario Fazio}
\affiliation{ICTP, Strada Costiera 11, 34151 Trieste, Italy}
\affiliation{NEST, Scuola Normale Superiore \& Istituto Nanoscienze-CNR, I-56126 Pisa, Italy}
\author{Cristiano Ciuti}
\affiliation{Universit\'{e} Paris Diderot, Sorbonne Paris Cit\'{e}, Laboratoire Mat\'{e}riaux et Ph\'{e}nom\`{e}nes Quantiques, CNRS-UMR 7162, 75013 Paris, France} 
\author{Iacopo Carusotto}
\affiliation{INO-CNR BEC Center and Dipartimento di Fisica, Universit\`a di Trento, I-38123 Povo, Italy}

\begin{abstract}
We introduce a novel frequency-dependent incoherent pump scheme with a square-shaped spectrum as a way to study strongly correlated photons in arrays of coupled nonlinear resonators. This scheme can be implemented via a reservoir of population-inverted two-level emitters with a broad distribution of transition frequencies. Our proposal is predicted to stabilize a non-equilibrium steady state sharing important features with a zero-temperature equilibrium state with a tunable chemical potential. We confirm the efficiency of our proposal for the Bose-Hubbard model by computing numerically the steady state for finite system sizes: first, we predict the occurrence of a sequence of incompressible Mott-Insulator-like states with arbitrary integer densities presenting strong robustness against tunneling and losses. Secondly, for stronger tunneling amplitudes or non-integer densities, the system enters a coherent regime analogous to the superfluid state. In addition to an overall agreement with the zero-temperature equilibrium state, exotic non-equilibrium processes leading to a finite entropy generation are pointed out in specific regions of parameter space. The equilibrium ground state is shown to be recovered by adding frequency-dependent losses. The promise of this improved scheme in view of quantum simulation of the zero temperature many-body physics is highlighted.%
\end{abstract}
\maketitle
\section{Introduction}
\label{sec:intro}
Over the last decade, technological developments in optical devices have allowed to engineer materials presenting strong photon-photon interactions~\cite{Carusotto_rev,Hartmann_rev,Noh}, and a growing interest has been devoted to the possibility of stabilizing strongly correlated photonic gases. The so-called photon blockade regime~\cite{Imamoglu}, in which photons behave as impenetrable particles, has been reached in various single-mode cavity platforms embedding atoms~\cite{Birnbaum}, superconducting qubits~\cite{Lang}, or quantum dots~\cite{Faraon,Atac}, as well as in Rydberg EIT atomic gases~\cite{Pritchard,Peyronel}. However, even though non-~\cite{Underwood,Hafezi_topo,Tanese} or weakly-interacting gases~\cite{Superfluid,Lai,Jacqmin} have been widely studied in spatially extended systems, scaling up strong non-linearities into large lattices still remains an open challenge~\cite{Fitzpatrick}.

Overcoming such an obstacle would be an essential step toward the stabilization of novel photonic phases \cite{Jin_davide1,Jin_davide2,Jin_alberto}, including Fractional Quantum Hall~\cite{Laughlin,Onur,Kapit} and Mott Insulator (MI) states~\cite{Fisher,Spielman}  in which the photon blockade prevents the onset of extended coherence. In particular, this latter state of photons has been predicted for Bose-Hubbard (BH)~\cite{Hartmann} and Jaynes-Cummings-Hubbard~\cite{Angekalis,Greentree} models in the limit of an isolated equilibrium photonic system, under the requirements of an integer density and low enough temperatures. 
In order to apply these results to experiments, one has however to keep in mind that the particle number is hardly conserved in realistic set-ups and heating effects cannot be neglected. It is thus essential to develop general schemes to tame and possibly exploit the intrinsic non-equilibrium nature of photonic systems. 

While, up to now, most studies focused on how to use a coherent drive to refill the photonic population~\cite{Carusotto_coherent,tomadin,Henriet,Biondi,Hoffman_Schmidt}, a few recent works~\cite{Gartner_Hartmann,Kapit,Lebreuilly,Ma_Simon} have suggested to employ a {\em frequency-dependent incoherent pump} to stabilize interesting incompressible photonic phases. The narrow bandpass frequency spectra (e.g. Lorentzian ones) considered in these works appear in fact well suited to observe Fractional Quantum Hall effects~\cite{Kapit} under the requirement of flat conduction bands~\cite{Kapit_Muller}, to stabilize strongly localized $n=1$ Mott insulator states~\cite{Lebreuilly,Ma_Simon}, as well as to perform some quantum error correction operations \cite{Kapit_Quantum}. However, they do not appear suitable to explore the dissipative photonic Bose-Hubbard model looking for non-equilibrium phase transitions driven by the usual interplay of hopping and interactions for which the hole spectral density typically has a significant linewidth. Namely, it was demonstrated in~\cite{Alberto} that even weak values of the hopping amplitude result in a proliferation of holes inside the Mott state, which then undergoes a transition toward a coherent superfluid-like phase with incommensurate density.

In order to study non-equilibrium counterparts of the well-known Mott-insulator to superfluid phase transition of equilibrium statistical mechanics, it is thus essential to develop new pumping schemes that allow to refill holes across a wide bandwidth. An interesting first step in this direction was discussed in~\cite{Hafezi} where a relatively complex protocol was proposed to implement the idea of a chemical potential for light in a circuit-QED platform. For suitably chosen equilibration rates with the engineered reservoir, photons may effectively thermalize to a statistical distribution with the desired thermodynamic parameters.

In this work we follow a different and potentially much simpler path of proposing the use of non-Markovian incoherent baths with tailored emission and loss spectra to stabilize the desired many-body states and, on a longer run, observe interesting phase transitions. A simple yet realistic implementation of a non-Markovian incoherent pump with a tailored emission spectrum can be obtained by using population-inverted two-level emitters with a broad distribution of transition frequencies. A similar scheme with a wide frequency distribution of absorbers and/or additional lossy cavities can be used to tailor the absorption spectrum.

As a first result, we show how a square-like emission spectrum allows to cool strongly correlated photonic systems toward ground-state-like steady states with a tunable effective chemical potential. We illustrate our proposal on the paradigmatic case of a one-dimensional (1D) BH model and we numerically show how it is possible to stabilize Mott-Insulator-like states with an arbitrary integer and fluctuationless photon density which are robust against tunneling and losses. For higher tunneling amplitudes or non-integer densities, our finite-size system exhibits a crossover towards a coherent state reminiscent of the Mott insulator to superfluid transition of equilibrium systems.

In addition to the overall agreement with the equilibrium physics, novel non-equilibrium processes leading to entropy generation and deviation from a zero temperature state are unveiled and characterized in some specific regions of the parameters space.
In order to overcome this entropic transition and be able to perform a full quantum simulation of the whole phase diagram of the Bose-Hubbard model, we extend the model by adding frequency-dependent losses. We anticipate and confirm numerically that in this way the steady-state fully overlaps with the zero-temperature equilibrium state for all choices of parameters.

The structure of the article is the following. In Sec.\ref{sec:model}, we introduce the physical model and in Sec.\ref{sec:exp_proposal} we discuss a possible experimental implementation. In Sec.\ref{sec:steady-state_equilibrium}, we discuss the equilibrium-like properties of the non-equilibrium steady-state in the presence of a frequency-dependent emission and we briefly review the single cavity physics. Numerical characterization of the insulator- and superfluid-like non-equilibrium steady-states of finite one-dimensional chains are presented in Sec.\ref{sec:numerical_results} first for an idealized set of parameters, then for realistic parameters inspired to state-of-the-art circuit-QED implementations. Intriguing non-equilibrium features leading to entropy generation in some regions of the parameter space are highlighted and explained in Sec.\ref{sec:non-equilibrium_features}. The performances of the extended model involving additional frequency-dependent losses are discussed in Sec.~\ref{sec:frequency-dependent_losses}. Conclusions are finally drawn in Sec.\ref{sec:conclusion}. In the Appendices, we provide more details on the emission spectrum (App.~\ref{app:analytical_spectrum}), on the derivation of the Redfield master equation (App.~\ref{app:projective}), on the possibility of pumping only a restricted number of sites and of recovering the square-shaped emission spectrum using a single temporally modulated emitter (App.~\ref{app:scheme_simplifications}). 

\section{The model}
\label{sec:model}
We consider a driven-dissipative model for strongly interacting photons in an array of $L$ coupled nonlinear cavities: 
\begin{equation}
\label{eq:BH}
H_{\rm ph}=\sum_{i=1}^{L}\left[\omega_{\rm{cav}} a_{i}^{\dagger}a_{i}+\frac{U}{2}a_{i}^{\dagger}a_{i}^{\dagger}a_{i}a_{i}\right]{-}\sum_{\avg{i,j}}Ja_{i}^{\dagger}a_{j},
\end{equation}
where $a_{i}$ ($a_i^\dagger$) are bosonic annihilation (creation) operators for photons in the $i$-th cavity. As usual, $J$ is the tunneling amplitude between neighboring cavities and $U$ is the on-site interacting energy.

\begin{figure}
\includegraphics[width=0.98\columnwidth,clip]{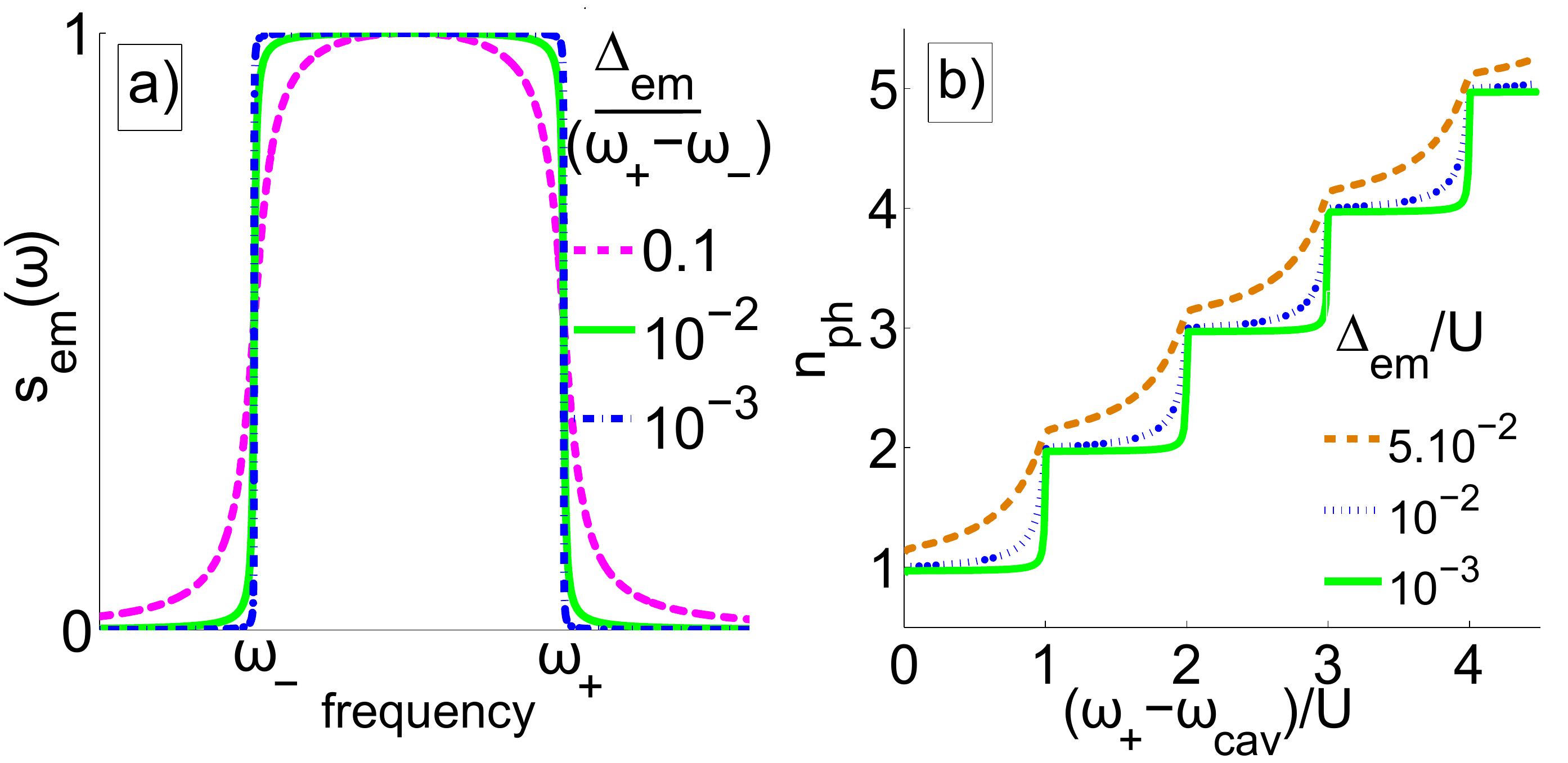}
\caption{\label{fig:square-spectrum+plateau} Panel a): Plot of the ``square-shaped'' emission spectrum ${s}_{\rm em}(\omega)$ defined in (\ref{eq:square_spectrum}) for various values of $\Delta_{\rm{em}}$. Panel b):  Average photon number $n_{\rm{ph}}$  as a function of $\mu=\omega_{+}-\omega_{\rm{cav}}$ (i.e. varying $\omega_{\rm{cav}}$) for a single site system with various values of $\Delta_{\rm{em}}$. Parameters of panel b): $\Gamma_{\rm{em}}^0/U=3.10^{-4}$, $\Gamma_{\rm{l}}/U= 10^{-5}$, $\omega_{-}/U=-40$.}
\end{figure}

In this work, we focus on the weakly-dissipative regime, in which photonic losses and emission processes are slow with respect to the bath memory time scales. The time-evolution of the density matrix $\rho$ then obeys the following Redfield master equation~\cite{Breuer} ($\hbar=1$):
\begin{equation}
\partial_{t}\rho(t) =  -i\left[H_{\rm ph},\rho(t)\right]+\mathcal{L}_{\rm{l}} \big[\rho(t)\big]+ \mathcal{L}_{\rm{em}}\big[\rho(t)\big],
\label{eq:photon_only}
\end{equation}
where $[ \cdot, \cdot]$ indicates as usual the commutator of two operators.

While losses are assumed to be Markovian and therefore modeled by a usual Lindblad term 
\begin{equation}
 \mathcal{L}_{\rm{l}}[\rho]=\frac{\Gamma_{\rm{l}}}{2}\sum_{i=1}^{L}\mathcal{D}[a_{i};\rho]
\end{equation} 
with $\mathcal{D}[\mathcal{O};\rho]=2\mathcal{O}\rho\mathcal{O}^\dagger-\mathcal{O}^\dagger \mathcal{O}\rho-\rho \mathcal{O}^\dagger \mathcal{O}$, the key novelty of this work is to use a frequency-dependent incoherent pump, so that the emission term
\begin{equation}
\label{eq:gain-non-markov}
\mathcal{L}_{\rm{em}}\big[\rho\big] = \frac{\Gamma_{\rm{em}}^{0}}{2}\sum_{i=1}^{L}\left[\tilde{a}_{i}^{\dagger}\rho a_{i}+a_{i}^{\dagger}\rho\tilde{a}_{i}-a_{i}\tilde{a}_{i}^{\dagger}\rho-\rho\tilde{a}_{i}a_{i}^{\dagger}\right]\!
\end{equation}
does not have a standard Lindblad form and involves modified lowering ($\tilde a_i$)
and raising ($\tilde a^\dagger_i\equiv[\tilde a_i]^\dagger$) operators:
\begin{equation}
\label{eq:special-operators}
\frac{\Gamma^{0}_{\rm{em}}}{2}\tilde{a}_{i} = \!\int_{0}^{\infty}d\tau\, \Gamma_{\rm{em}}(\tau)a_{i}(-\tau).
\end{equation}
Here, the kernel 
\begin{equation}
\label{eq:memory-kernel}
\Gamma_{\rm{em}}(\tau)=\theta(\tau)\int \frac{d\omega}{2\pi}\mathcal{S}_{\rm{em}}(\omega)e^{-i\omega\tau}
\end{equation}
 takes into account the reservoir emission  spectrum $\mathcal{S}_{\rm{em}}(\omega)$, while the $a_i(t)$ operators are defined in the interaction picture with respect to the photonic Hamiltonian, $a_{i}(\tau)=e^{iH_{\rm ph}\tau}\,a_{i}\,e^{-iH_{\rm ph}\tau}.$
Thus, considering two eigenstates $\ket f$ (resp. $\ket{f'}$) of the photonic Hamiltonian with $N$ (resp. $N+1)$ photons and energy $\omega_{f}$ (resp. $\omega_{f'}$), the matrix element of the modified jump operators equals
\begin{equation}
\label{eq:special-operators-frequency}
\bra{f}\tilde{a}_i\ket{f'} =\frac{2}{\Gamma_{\rm{em}}^0} \Gamma_{\rm{em}}(\omega_{f'f}) \bra{f}a_i\ket {f'},
\end{equation}
with $\omega_{f'f}=\omega_{f'}-\omega_{f}$ and 
\begin{equation}
\Gamma_{\rm{em}}(\omega) = \frac{1}{2} \mathcal{S}_{\rm{em}}(\omega)-i\delta_{l}(\omega)
\end{equation}
is the Fourier transform of the memory kernel $\Gamma_{\rm{em}}(\tau)$.
While the magnitude of the Lamb-shift $\delta_{l}(\omega)$ stemming from the imaginary part of $\Gamma_{\rm{em}}(\omega)$ is typically small as compared to the emission linewidth (see  App.~\ref{app:analytical_spectrum}) and thus does not bring important physical effects, the real part $\mathcal{S}_{\rm{em}}(\omega)/2$ is physically essential as it provides the frequency-dependent emission rate. Further extension of the model including non-Markovian losses will be discussed in Sec.\ref{sec:frequency-dependent_losses}.

The physics  and the phase diagram of this driven-dissipative model critically depend on the specific choice of the emission spectrum. In contrast with our previous work~\cite{Lebreuilly} in which the emission spectrum was Lorentzian, we will focus here on the study of a ``square-shaped" spectrum $\mathcal{S}_{\rm{em}}(\omega)= s_{\rm em}(\omega) \Gamma^{0}_{\rm em}$, where 
\begin{equation}
s_{\rm{em}}(\omega)=\mathcal{N}\,\int_{\omega_{-}}^{\omega_{+}} d\omega' \frac{\Delta_{\rm{em}}/2}{(\omega-\omega')^2+(\Delta_{\rm{em}}/2)^2}
\label{eq:square_spectrum}
\end{equation}
is shown in Fig.\ref{fig:square-spectrum+plateau} a) and the normalization constant $\mathcal{N}$ is set such that $s_{em}\left(\frac{\omega_++\omega_-}{2}\right)=1$.
From the figure, one sees that $\mathcal{S}_{\rm{em}}(\omega)$ maintains an almost constant value $\Gamma_{\rm{em}}^0$ all over a frequency domain $[\omega_- ,\omega_+]$, and decays smoothly with a power law outside this interval over a frequency scale $\Delta_{\rm{em}}\ll \omega_{+}-\omega_{-}$ (more details on analytical expressions for the emission spectral properties can be found in App.~\ref{app:analytical_spectrum}) .

The lower cutoff does not play any role in our proposal and may be set to a very  far red-detuned frequency, $\omega_{\rm{cav}}-\omega_{-}\gg U,J\geq 0$. Experimentally, a value of $\omega_{\rm{cav}}-\omega_{-}$ on the order of a few times  $U$ is typically enough. On the other hand, a key role is played by the upper cutoff $\omega_{+}$: in Sec.\ref{sec:steady-state_equilibrium}, we will show that the detuning $\omega_+-\omega_{\rm{cav}}$ sets a sort of chemical potential for photons. Finally, the weak dissipation condition that we assumed at the beginning of this section translates into $\Gamma_{\rm{l}},\,\Gamma_{\rm{em}}^{0}\ll \Delta_{\rm{em}},\, (\omega_{+}-\omega_{-})$.

\section{Experimental proposal}
\label{sec:exp_proposal}

\subsection{Ideal configuration}
\label{sec:ideal_conf}
To engineer the non-Markovian pump introduced in Eq.~(\ref{eq:gain-non-markov}), we propose to insert a large number $N_{\rm{at}}\gg 1$ of two-level emitters into each cavity. Their evolution and their coupling to the cavity field are described by Hamiltonian terms of the form
\begin{eqnarray}
H_{\rm{at}}&=&\sum_{i=1}^{L}\sum_{n=1}^{N_{\rm{at}}}\omega_{\rm{at}}^{(n)}\sigma_{i}^{+(n)}\sigma_{i}^{-(n)}\\
H_{I}&=&\Omega_{R}\sum_{i=1}^{L}\sum_{n=1}^{N_{\rm{at}}}(a_{i}^{\dagger}\sigma_{i}^{-(n)}+{\rm h.c.}),
\end{eqnarray}
where $\sigma_{i}^{-(n)}$ ($\sigma_{i}^{+(n)}$) are the lowering (raising) operators for the two-level $n$-th emitter in the $i$-th cavity and $\Omega_R$ is the single-emitter Rabi frequency. The transition frequencies $\omega_{\rm{at}}^{(n)}$ of the different emitters are assumed to be uniformly distributed over the interval $[\omega_- ,\omega_+]$. 

Each emitter is incoherently pumped in the excited state at a rate $\Gamma_{\rm{p}}$, which is modeled by the Lindblad term 
\begin{equation}
\mathcal{L}_{\rm{p,at}}[\rho_{\rm{tot}}]=\frac{\Gamma_{\rm{p}}}{2}\sum_{i=1}^{L}\sum_{n=1}^{N_{\rm{at}}}\mathcal{D}[\sigma_{i}^{+(n)};\rho_{\rm{tot}}]
\end{equation}
so that the total (cavity+emitters) density matrix $\rho_{\rm{tot}}$ obeys the master equation:
\begin{eqnarray}
\partial_{t}\rho_{\rm{tot}}(t)&=&-i\com{H_{\rm ph}+H_{\rm at}+H_{I}}{\rho_{\rm{tot}}(t)}\\
&&\qquad+\mathcal{L}_{\rm{l}}\big[\rho_{\rm{tot}}(t)\big]+\mathcal{L}_{\rm{p,at}}\big[\rho_{\rm{tot}}(t)\big].\nonumber
\end{eqnarray}
For sufficiently strong pump rate $\Gamma_{\rm{p}}$, the pump induces an almost perfect inversion of population in the emitters. As a result, these undergo irreversible cycles in which they are immediately re-pumped after emitting a photon in the cavity and reabsorption processes are suppressed. Such a pumping can be implemented for instance by coherently driving the emitter into a third level, from which it quickly decays towards the excited state of the active transition as often done in practical laser devices and discussed in~\cite{Ma_Simon}.

Due to the broadening induced by the pump, each emitter displays a Lorentzian emission spectrum of linewidth $\Gamma_{\rm{p}}$
\begin{equation}
\label{eq:Lorentzian_contribution}
\mathcal{S}_{\rm{em}}^{(single)}(\omega)=\Gamma_{\rm{em}}^{\rm{at}}\frac{(\Gamma_{\rm{p}}/2)^2}{(\omega-\omega_{\rm{at}})^2+(\Gamma_{\rm{p}}/2)^2}
\end{equation}
with $\Gamma_{\rm{em}}^{(at)}=\frac{4\Omega_{R}^2}{\Gamma_{\rm{p}}}$ (see App.~\ref{app:projective}). Integration over the contribution of all emitters across their uniform frequency distribution $[\omega_-,\omega_+]$ yields the desired square-shaped spectrum of Eq.~(\ref{eq:square_spectrum}) and Fig.~\ref{fig:square-spectrum+plateau} a).

Technically speaking, under the constraints $\sqrt{N_{\rm{at}}}\Omega_{R},\Gamma_{\rm{l}}\ll\Gamma_{\rm{p}}$, we can use projective methods~\cite{Breuer} to trace out the emitter degrees of freedom~(see App.~\ref{app:projective} for a sketch a the method and \cite{Lebreuilly} for a full derivation) and write a closed master equation for the photonic density matrix in the form of Eq.~(\ref{eq:photon_only}) with $\mathcal{S}_{\rm{em}}(\omega)$ given by Eq.~(\ref{eq:square_spectrum}), with an edge width equal to $\Delta_{\rm{em}}=\Gamma_{\rm{p}}$. For $\Gamma_{\text{p}}=\Delta_{\text{em}}\ll\omega_{+}-\omega_{-}$, we obtain for the maximum emission rate $\Gamma_{\rm{em}}^0=2\pi N_{\rm{at}}\Omega_{R}^2 /(\omega_{+}-\omega_{-})$.

\subsection{Possible simplification strategies}
\label{subsec:simplification_scheme}
At a first glance, the physical implementation of the proposed scheme may appear as a quite challenging task, as it involves coupling a large number of different emitters to each resonator. In the following part of this section, we are going to explain how this scheme may be simplified and made accessible to state-of-the-art technology.

\subsubsection{Pumping a few sites only}

As a first idea, following a suggestion of \cite{Ma_Simon}, we may argue that a pumping mechanism restricted to one or two sites only is sufficient to stabilize the same steady-state that one would obtain if emitters were present on all sites. While a more complete numerical validation of the principle under different boundary conditions can be found in App.~\ref{app:pumping_a_few_sites}, the following physical arguments already support our claim. 

When a photon is lost starting from a state $\ket{f}$, a sort of ''hole`` is created in the fluid. Due to tunneling, this hole can travel along the chain at a significant group velocity that typically scales as $v_g\sim J\gg \Gamma_{\rm{l}},\,\Gamma_{\rm{em}}^0$ and is thus able to expand over a large number of sites before undergoing decoherence. Because of this delocalization effect, many sites (not only the one where the initial loss process took place) feel the presence of the hole and are able to replenish the original many-body state $\ket{f}$ by injecting a new photon. 
Of course, the single-site emission rate has to be correspondingly increased to maintain the same total emission power. As pointed out in \cite{Ma_Simon}, in case of large lattices, hole excitations (in particular low-momentum ones with slow group velocities) might not have the time to travel and reach the emitting site before suffering from additional dissipative processes. In this case, it is enough to introduce many regularly spaced emitting sites in the bulk of the chain.

Finally, attention must be paid so to avoid the emitter being located at the node of the wave function of some hole states, which would block the re-emission of a new photon. Indeed, due to reflection symmetry, generated hole wave packets possess a symmetric momentum distribution and therefore must be seen as a superposition of cosine-like standing waves. For the most relevant experimental configuration of open boundary conditions, this issue can be avoided in a simple manner by setting the emitting site at one of the chain extremities, where the nodes can not be located. For a periodic chain where the location of those nodes is not fixed due to the absence of edges, embedding emitters in two neighboring sites is enough to have all hole states quickly replenished.

\subsubsection{Temporally-modulated emitter frequency}
\label{sec:temporal_modulation}
Along different lines, a dramatic reduction of the number of required emitters can be obtained by making a single emitter to mimic the effect of a square spectrum. In the original proposal presented in Sec.\ref{sec:ideal_conf}, each single emitter provides a Lorentzian contribution (Eq.~(\ref{eq:Lorentzian_contribution})) to the emission spectrum and the square spectrum of Eq.~(\ref{eq:square_spectrum}) is recovered upon integration over a uniform distributed of emitter frequencies within the interval $[\omega_{-},\omega_{+}]$. 

The goal of this subsection is to suggest how a wide distribution of emitters can be imitated by temporally modulating the transition frequency of a single emitter. In order to get a uniform distribution, one needs a constant modulation speed $v_\omega=|\frac{d\omega_{\rm{at}}}{dt}|$. Under suitable conditions described below, the resulting time-averaged emission spectrum then has the desired shape
\begin{eqnarray}
\mathcal{S}_{\rm{em}}^{(av)}(\omega)&=&\frac{\Gamma_{\rm{em}}^{\rm{at}}}{T}\int_t^{t+T}\!\!\!\!\!\!\!\! dt\,\frac{(\Gamma_{\rm{p}}/2)^2}{(\omega-\omega_{\rm{at}}(t))^2+(\Gamma_{\rm{p}}/2)^2}\\
&=&\frac{\Gamma_{\rm{em}}^{\rm{at}}}{\omega_{+}-\omega_-}\int_{\omega_{-}}^{\omega_{+}} \!\!\!\!\!d\tilde{\omega}\,\frac{(\Gamma_{\rm{p}}/2)^2}{(\omega-\tilde{\omega})^2+(\Gamma_{\rm{p}}/2)^2},\nonumber
\end{eqnarray}
where  $T=\frac{(\omega_+-\omega_{-})}{v_\omega}$ is the frequency modulation half-period. A similar idea was experimentally implemented in \cite{Hoffman_Schmidt} to obtain a square spectrum field by modulating a classical source in time. This technique allowed to spectrally probe the different photonic levels of a single mode cavity coupled to a far-off-resonance emitter, and thus to demonstrate a dispersive blockade effect. 

While a full discussion of this method is postponed to a further publication, one can already see on physical grounds that in order to avoid to avoid spurious effects, several conditions must be met. First, $v_{\omega}$ should be fast enough for photons not be lost within a modulation half-period $T$, which imposes that $\frac{1}{T}=\frac{v_{\omega}}{\omega_+-\omega_{-}}\gg \Gamma_{\rm{l}}$. If this condition is not satisfied, the scheme fails to stabilize a quasi time-independent steady-state and the system keeps performing wide oscillations. 

As a second requirement, $v_{\omega}$ should be slow enough that well-defined edges are maintained at the extremes of the spectrum and uncontrolled heating effects are avoided. Provided $v_{\omega}\ll\Gamma_{\rm{p}}^2$, the resulting frequency-dependent emission is expected to converge toward the exact square spectrum of Fig.~\ref{fig:square-spectrum+plateau} a) with edges possessing a width $\Delta_{\rm{em}}=\Gamma_{\rm{p}}$. Otherwise, one expects that $1/\sqrt{v_{\omega}}$ becomes then the dominant limiting time scale in the memory kernel of Eq.~(\ref{eq:memory-kernel}) and the effective edge linewidth increases as $\Delta_{\rm{em}}^{\rm{eff}}\propto \sqrt{v_{\omega}}$. As we shall see in Sec.~\ref{sec:steady-state_equilibrium}, this additional broadening can be tolerated as long as one remains in the $\Delta_{\rm{em}}^{\rm{eff}}\ll U$ regime for frequency-selectivity.

These two constraints can be simultaneously satisfied for weak enough losses. As we will see in Sec.~\ref{sec:steady-state_equilibrium}, $\omega_{+}-\omega_{-}$ should be at least equal to a few times the interaction strength $U$, and $\Gamma_{\rm{p}}=\Delta_{\rm{em}}$ is a tunable parameter which will have to verify $\Delta_{\rm{em}}/U\ll 1$: one concludes that a very small $\Gamma_{\rm{l}}/U\lll 1$, allows to simultaneously satisfy both conditions. Some preliminary numerical checks validating this conclusion as well as our intuition on how to optimize the performance of this scheme are presented in App.\ref{app:temporally_modulated} for a very simplified single-cavity model.

For realistic parameters a compromise between the two opposite constraints must be found. In that prospect, a good strategy may be to use several emitters spanning different sub-intervals of the spectral range $[\omega_{-},\omega_{+}]$: in this way, the modulation speed $|\frac{d\omega_{\rm{at}}}{dt}|$ required  to cover the whole interval $[\omega_{-},\omega_{+}]$ within the finite photon lifetime $1/\Gamma_{\rm{l}}$ would in fact be reduced, which would help fulfilling the two constraints.

\begin{figure*}
\includegraphics[width=0.98\textwidth,clip]{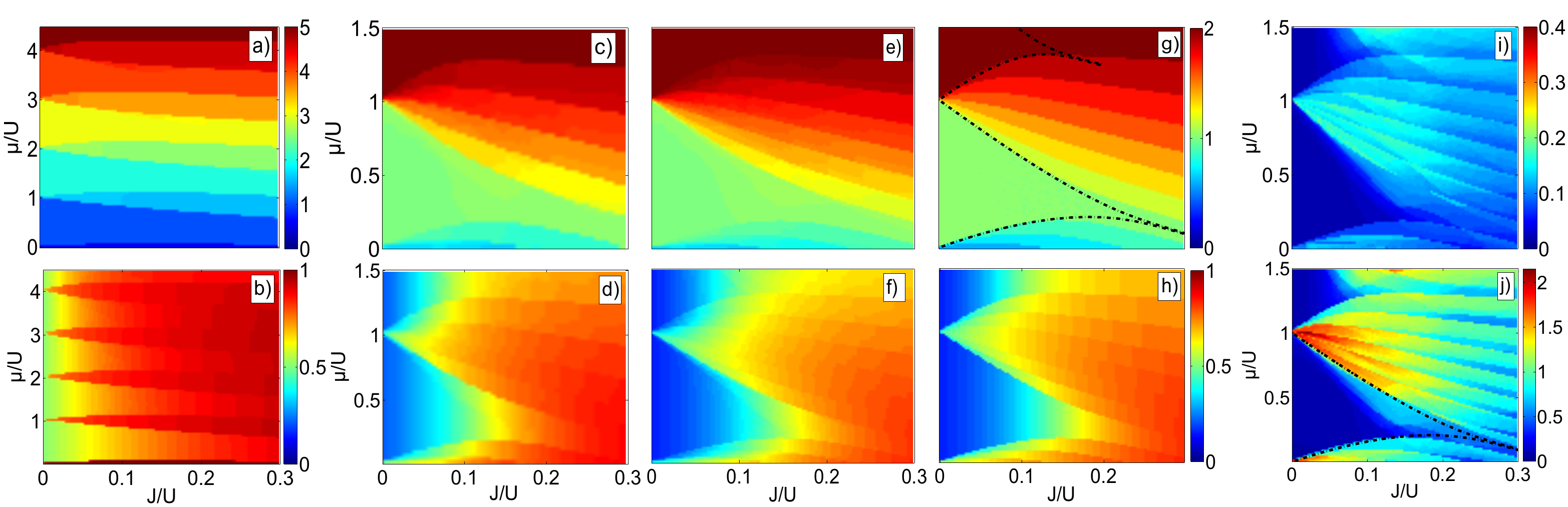}\vspace{-4mm}
\caption{\label{fig:phase_diagram}Steady-state properties for a limit choice of parameters. Panels a),c),e) [resp. panels b),d),f)] Average steady-state photon number per site $n_{\rm{ph}}$ (resp. condensed fraction $x_{\rm{BEC}}$), for $L=2,5,7$ respectively. Panel g) [resp. panel h)]: Average $n_{\rm{ph}}$  (resp. $x_{\rm{BEC}}$) in a $T=0$ equilibrium system, for $L=7$. Panel i) [resp. panel j)]: steady-state particle number relative fluctuations $\Delta n$  [resp. entropy $S=-\langle \text{ln}(\rho_{\infty})\rangle$], for $L=7$. In panel g) and j)  dash dotted black lines indicate the MPS $T=0$ prediction for the first Mott lobes. Parameters used in all panels (except g,h): $\Gamma_{\rm{l}}/\Gamma_{\rm{em}}^0=10^{-3}$,  $U/\Delta_{\rm{em}}=10^6$, $\Gamma_{\rm{em}}^0/\Delta_{\rm{em}}=10^{-2}$, $\omega_{+}=0$, $\omega_{-}/\Delta_{\rm{em}}=-4\,\times10^7$. Cutoff in particle number per site $N_{\rm{max}}=6$ [Panels a),b)] and $N_{\rm{max}}=3$ [panels c)-j)]
}
\end{figure*}

\section{Steady-state equilibrium-like properties}
\label{sec:steady-state_equilibrium}
In a single cavity geometry, the scheme introduced in this work allows to stabilize pure Fock states with arbitrary photon number: indeed, in Fig.~\ref{fig:square-spectrum+plateau} b) we observe a plateau structure with successive jumps  between integer values of the steady-state photon number, with a smooth (resp. sharp and discontinuous)  transition for  $\Delta_{\rm{em}}/U\gtrsim 1$ (resp. $\Delta_{\rm{em}}/U\ll 1$) between the various steps. The physical quantity $\omega_{+}-\omega_{\rm{cav}}$ manifestly plays a similar role as the chemical potential in equilibrium physics~\cite{Fisher} but with a mechanism differing from~\cite{Hafezi}.

Pushing the analogy with equilibrium forward, we find that the specific shape of the pump spectrum 
[Fig.~\ref{fig:square-spectrum+plateau} a)] allows to drive large many-cavity systems toward a steady state closely related to a $T=0$ state (and thus to overcome the fragility against tunneling pointed out in~\cite{Lebreuilly} for Lorentzian pumps). To see this, let us set a strong emission at resonance $\Gamma_{\rm{em}}^0\gg \Gamma_{\rm{l}}$, while maintaining a sharp cut-off at the edges of the spectrum $\Delta_{\rm{em}}\ll U$, in such a way to strongly favor (resp. block) $f\rightarrow f'$ transitions between eigenstates of the photonic Hamiltonian (with $N$ and $N+1$ photons) verifying $\omega_{f'f}\leq\omega_+$ (resp. $\omega_{f'f}\geq\omega_+$). Under those constraints, the transition rates follow the condition
\begin{eqnarray*}
\label{eq:detailed-balance}
\frac{\mathcal{T}_{f\to f'}}{\mathcal{T}_{f'\to f}}\simeq \frac{\Gamma^0_{\rm{em}}}{\Gamma_{\rm{l}}}\theta(\omega_{+}-\omega_{f'f}) \quad\left\lbrace\begin{array}{ll}
\ggg 1& \text{ if}\,\, \omega_{f'f}<\omega_{+} \\
\lll 1& \text{ if} \,\,\omega_{f'f}>\omega_{+}
\end{array}\right. ,
\end{eqnarray*}
that closely resembles  a $T=0$ detailed-balance relation
\begin{equation}
\left. \frac{\mathcal{T}_{f'\to f}}{\mathcal{T}_{f\to f'}}\right|_{\rm eq}=e^{\beta(\omega_{+}-\omega_{f'f})}
\label{eq:detailed_balance2}
\end{equation}
with $\beta\to +\infty$. One may thus expect the many-body steady state to be very close to the ground state $\ket{\rm GS}$ of the rotating frame Hamiltonian 
\begin{eqnarray}
\label{eq:effective_hamiltonian}
H_{\rm eff}&=&H_{\rm ph}-\omega_{+}N\\
&=&\sum_{i=1}^{L}\left[-\mu a_{i}^{\dagger}a_{i}+\frac{U}{2}a_{i}^{\dagger}a_{i}^{\dagger}a_{i}a_{i}\right]{-}\sum_{\avg{i,j}}Ja_{i}^{\dagger}a_{j},\nonumber
\end{eqnarray}
i.e., a $T=0$ state with chemical potential $\mu=\omega_{+}-\omega_{\rm{cav}}$. 

In the next sections, we will show that this agreement with equilibrium physics is generally robust for most choices of system parameters, but subtle signatures of the non-equilibrium condition and of the deviation from the detailed balance condition (\ref{eq:detailed_balance2})  can appear in some specific regions. These new non-equilibrium features will be discussed at length in Sec.\ref{sec:non-equilibrium_features}. A way to suppress them by adding extra frequency-dependent losses is then introduced and characterized in Sec.~\ref{sec:frequency-dependent_losses} in view of quantum simulation applications.

\section{Numerical results for finite periodic chains}
\label{sec:numerical_results}

As the sophisticated numerical techniques used in~\cite{Jin_alberto,Alberto} are not straightforwardly applicable to non-Markovian problems, we had to base our study on a direct numerical calculation of the steady-state density matrix $\rho_\infty\equiv\rho(t \to +\infty)$ by looking for a zero of the Redfield super-operator on the right-hand side of Eq.~(\ref{eq:photon_only})  for mesoscopic one-dimensional chains. While a complete study of larger systems in possibly higher dimensionality is postponed to future work addressing e.g. the critical properties of possible phase transitions, our approach turned out to be sufficient to anticipate and understand the behaviour of experimentally relevant systems.

For system sizes going up to $L=5$ sites, a complete numerical calculation was possible. Above 5 sites, we had to perform the secular approximation and discard fast oscillating terms in the master equation: for very weak dissipation and in absence of relevant degeneracies of the photonic Hamiltonian, the diagonal terms of the density matrix in the Hamiltonian eigenbasis are in fact not coupled to off-diagonal terms, and the latter can be neglected when computing the steady state. As this approximation is generally accurate for weak dissipation but may be problematic in the presence of degeneracies, we have numerically checked on chains of $L=3, 4, 5$ sites that it indeed gives indistinguishable results from the exact solution for small systems sizes.

In order to facilitate the reader, we start our discussion in Sec.\ref{sec:ideal} from a limit case of parameters for which the physics is most transparent [Fig.~\ref{fig:phase_diagram}]. As a second step, in Sec.\ref{sec:sota} we will then assess the robustness and actual observability of our predictions by considering parameters inspired to state-of-the-art experimental devices [Fig.~\ref{fig:phase _diagram_realistic}].

\subsection{Idealized parameters}
\label{sec:ideal}

In order to present the physics in a cleanest way,  we first discuss the occurrence of the insulator-like state and its transition towards a superfluid-like state for an idealized set of parameters where the loss rate $\Gamma_{\rm l}$ is extremely small as compared to the interaction energy $U$. This allows to keep all other parameters well spaced in magnitude and largely satisfy the inequalities.  Calculations showing the robustness of our conclusions for realistic parameters of state-of-the-art circuit-QED devices are presented in the next subsection.

The steady-state photon density $n_{\rm{ph}}=\langle N\rangle/L$ and the Bose-condensed fraction $x_{\text{BEC}}=\langle n_{k=0}\rangle/\langle N\rangle$ (where $\langle O\rangle\equiv{\rm Tr}(O\rho_{\infty})$) are given in Fig.~\ref{fig:phase_diagram} [panels a)-f)] for several system sizes $L$, and are compared to the $T=0$ equilibrium predictions for $L=7$ sites [panels g), h)]. Even though one does not expect a true BEC for an infinite 1D chain~\cite{Mermin_Wagner}, still $x_{\text{BEC}}$ provides physical insight on the long-range coherence properties of our finite-size system.

Apart from the presence of small corrections that will be discussed below, the qualitative agreement between the observables calculated for the driven-dissipative steady state and the $T=0$ prediction of the equilibrium BH model is very good: first, we observe for increasing $\mu$ a series of insulating-like regions with successive integer values of the density $n_{\rm{ph}}$ and a small $x_{\rm{BEC}}$.  Within these regions, the photonic density does not depend on the Hamiltonian parameters $\omega_{\rm{cav}}$ and $J$, and fluctuations in the total photon number are suppressed to {\footnotesize $\Delta n\equiv\sqrt{\langle N^2\rangle-\langle N\rangle^2}/\langle N \rangle\simeq 10^{-2}$}  [Fig.~\ref{fig:phase_diagram} i)]: this is a sort of non-equilibrium form of incompressibility.

These insulating regions closely follow the shape of the phase boundary [panel g)] predicted for a $T=0$ equilibrium 1D system (the so-called Mott lobes \cite{Fisher,Ejima}), that we obtained by means of matrix-product-states (MPS) simulations with $L=200$ sites (see \cite{Ejima} for details on the approach): the agreement for the first lobe is excellent, while the deviations for the second lobe are due to a numerical cutoff  in the maximum particle number per site $N_{\rm{max}}=3$ used in the steady-state calculation. 

Secondly, the insulating regions are separated by coherent regions with non-integer density, reminiscent of the equilibrium superfluid phase where excess particles/holes do not suffer from the photon blockade and can delocalize via tunneling: the condensed fraction is important and eventually reaches the maximal value $x_{\rm{BEC}}=1$ at high $J$, indicating a full coherence over the finite system.

\subsection{Realistic parameters}
\label{sec:sota}

While Fig.\ref{fig:phase_diagram} focused on a limit case of parameters in order to validate the theoretical viability of our approach, Fig.~\ref{fig:phase _diagram_realistic} confirms the actual feasibility of our proposal and the overall robustness of our predictions for state-of-the-art parameters in circuit-QED systems~\cite{Ma_Simon,Rigetti}. 

The main consequence of the finite ratios $\Delta_{\rm{em}}/U$ and $\Gamma_{\rm{em}}^0/\Gamma_{\rm{l}}$ is in fact a weak but appreciable value of  particle number fluctuations, and could be seen as the non-equilibrium counterpart of the effect of a finite temperature $T_{\rm{eff}}$. For a low-T equilibrium state with $\{J/U= 0,\mu/U=1/2\}$ (which is the point of the first lobe with highest energy gap and thus predictably at low temperature the one with the weakest fluctuations) and restricting the partition function $Z_{th}\simeq 1+e^{\beta\mu}+e^{\beta(2\mu-U)}$ to the most relevant Fock states $N=0,1,2$, one finds that the particle number fluctuations can be connected to the temperature through the following relation
\begin{equation}
kT=\frac{U}{2}\frac{1}{\text{ln}\left(\frac{2}{\Delta n^2}\right)}.
\end{equation}
A rough estimate for a sort of effective temperature for our non-equilibrium system can then be extracted by inserting in this formula the value $\Delta n \sim 0.13$ found in Fig.~\ref{fig:phase _diagram_realistic} c). This gives a quite low value $T_{\rm{eff}}\simeq 0.1\, U$. The steady-state effective temperature should be even lower away from $\{J/U= 0,\mu/U=1/2\}$ for the specific parameters choice of Fig.~\ref{fig:phase _diagram_realistic} c), since the resulting fluctuations are rather independent from tunneling and only change by a factor $\sim1.5$ across the transition line (in contrast to the equilibrium case where fluctuations dramatically increase when the many-body gap closes). Based on the low value of $T_{\rm{eff}}$, one can thus expect that it will be possible to catch the effect of quantum fluctuations (or crossover) for current state-of-the-art parameters, at least on some intermediate length scale.

\begin{figure*}
\includegraphics[width=0.9\textwidth,clip]{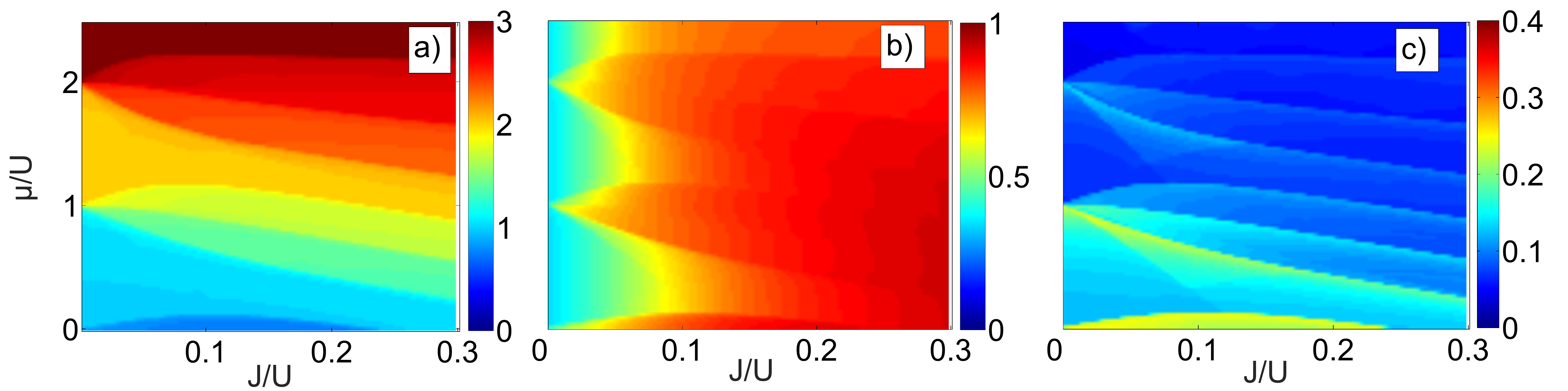}\hspace{-1mm}
\caption{\label{fig:phase _diagram_realistic} Steady-state properties for $L=3$, for state-of-the-art parameters in circuit QED. Panel a): average photon number per site $n_{\rm{ph}}$. Panel b): condensed fraction $x_{\rm{BEC}}$. Panel c):  relative fluctuations of the total particle number  $\Delta n $. Parameters inspired from circuit-QED systems~\cite{Ma_Simon,Rigetti}: $U=200\times 2\pi  \text{MHz}$, $\Delta_{\rm{em}}=0.5\times 2\pi \text{MHz}$, $\Gamma_{\rm{em}}^0=30\times 2\pi\text{kHz}$, $\Gamma_{\rm{l}}= 1\times 2\pi \text{kHz}$.  In order to be able to correctly see the higher lobes, we had to increase the maximum allowed number of particles per site to $N_{\rm{max}}=4$, and correspondingly to reduce the system size to $L=3$.}
\end{figure*}

\begin{figure}
\includegraphics[width=1\columnwidth,clip]{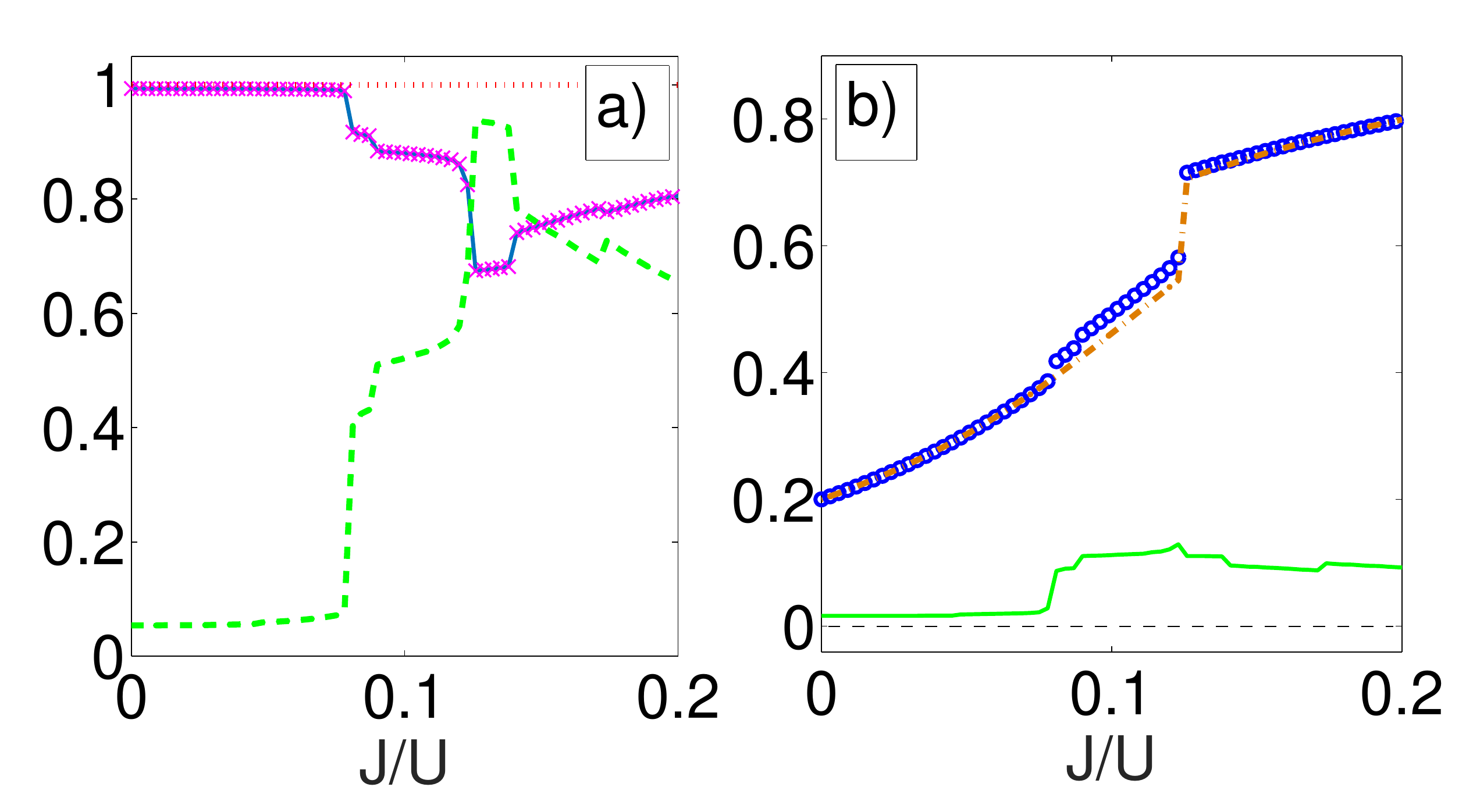}\vspace{-3mm}
\caption{\label{fig:steady_state_statistical} Steady-state statistical properties for $L=5$ at fixed $\mu=0.55\, U$. Panel a):  fidelity $\mathcal{F}$ between the steady state and the Hamiltonian ground state (blue solid line), occupancy $\pi_0$ of the most populated state $\ket{\psi}_{+}$ (purple crosses), overlap $|\langle\psi_+|GS\rangle|^2$ between the ground state and the most populated state  (red dot line), entropy $S$ (green dash line). Panel b): number fluctuations $\Delta n$ (green solid line) and condensed fraction (blue circles), compared to the $T=0$ equilibrium value  (orange dash-dot line). Same parameters as in Fig.~\ref{fig:phase_diagram}. In order to be able to perform exact diagonalization of the Liouvillian and avoid using the secular approximation, we had to choose a smaller system size $L=5$ as compared to Fig.~\ref{fig:steady_state_statistical}. }
\end{figure}

\subsection{Finite-size effects}
\label{sec:finite-size}

Even though there is a quite good overall agreement of the non-equilibrium calculations to the well-known physics of the equilibrium system in the thermodynamic limit \cite{Fisher,Ejima}, a careful observer can still notice in Fig.~\ref{fig:phase_diagram} some significant discrepancies, in particular with the $T=0$ prediction for the phase boundary [dash dotted line in Fig.~\ref{fig:phase_diagram} g)]. As a first step, it is therefore important to first assess which features are likely to be finite-size effects and which ones might instead signal some new physics.

The most prominent such features are that in Fig.~\ref{fig:phase_diagram} [b), d), f)] the insulating regions do not close completely to form lobes but rather end with a stripe, $x_{\rm{BEC}}$ is not exactly zero even at very weak $J$, and all observables present a smooth crossover for increasing tunneling instead of a sharp transition. Comparing these panels, one notices that both the width of the stripes and the condensed fraction inside the insulating region decreases as $1/L$ for increasing system sizes. The fact that a similar behaviour is found in the finite size equilibrium plot of Fig.~\ref{fig:phase_diagram} g) is therefore a strong indication of the finite-size origin of this effect. 

If one could take the infinite system size limit, a natural expectation would be that these discrepancies should in fact disappear, recovering clean Mott lobes surrounded by a superfluid (and possibly also Bose-condensed depending on the dimensionality) phase. However, as we are going to discuss in the next Section, a more careful analysis allows to unveil another kind of deviations, which signals a much richer non-equilibrium phenomenology.

\section{Non-equilibrium features}
\label{sec:non-equilibrium_features}

The most remarkable such feature is highlighted in the plots of the particle number fluctuations and of the entropy shown in Fig.~\ref{fig:phase_diagram}(i,j): While in most parts of the insulating region the steady state presents an almost vanishing entropy  $S=-\langle \text{ln}(\rho_{\infty})\rangle$ [Fig.~\ref{fig:phase_diagram} j)] and can thus be well approximated by a pure quantum state  (as for a $T=0$ equilibrium state), this is not the case in some regimes of parameters in the vicinity of the transition line where the entropy $S$ acquires a significant positive value slightly before the jump in the particle number and in the condensate fraction at the equilibrium superfluid transition. 

The present section is dedicated to the characterization of this novel entropic transition and to the description of the non-equilibrium mechanisms underlying it. Since this phenomenon could be seen as an hindrance in the prospect of quantum simulating the ground-state of the Bose-Hubbard Hamiltonian, in Sec.~\ref{sec:frequency-dependent_losses} we will put forward a further extension of the optical scheme that is able to remove this deviation from equilibrium.

To quantitatively characterize these non-equilibrium features, we looked at the fidelity $\mathcal{F}=\bra{\rm GS}\rho_{\infty}\ket{\rm GS}$ between the steady state $\rho_{\infty}$ and the Hamiltonian ground state $\ket{\rm GS}$ and at the steady-state occupancy $\pi_{0}$ of the most populated state $\ket{\psi_{+}}$~\footnote{To avoid possible artifacts such as the preferential choice of the steady-state eigenbasis, all these physical quantities were computed for $L=5$ by exact steady-state calculation without using the secular approximation.}. 

As one can see in Fig.~\ref{fig:steady_state_statistical}a), the transition takes the form of a discontinuous jump} in entropy from a $99\%$ pure quantum state toward a statistical mixture above some critical $J_c$, located within an insulating region at a small but finite distance from the equilibrium transition line [Fig.~\ref{fig:phase_diagram} j)]. Note that this jump is present even for finite sizes $L$, and just gets smoother for the state-of-the-art parameters of Fig.~\ref{fig:phase _diagram_realistic}. In the pure region $\mathcal{F}=\pi_0\simeq 0.993$ are close to unity, indicating that  $\rho_{\infty}$ can be well approximated by the pure state $\ket{\rm GS}\bra{\rm GS}$  (not the case in the entropic region). In both regions, $|\langle\psi_+|GS\rangle|^2=1$ (within machine precision),  $\mathcal{F}$ and $\pi_{0}$ take identical values, so the most populated state $\ket{\psi_+}$ is precisely equal to the Hamiltonian ground state at any value of $J$ (at least for the precise value of $\mu$ in Fig.~\ref{fig:steady_state_statistical} a)). Looking at the observable $x_{\rm{BEC}}$ [Fig.~\ref{fig:steady_state_statistical} b)], the pure  (resp. entropic) region is characterized by negligible (resp. small) deviations from equilibrium, and very weak fluctuations $\Delta n\simeq 0.016$ (resp. non-zero $\Delta n\simeq 0.13$).

\begin{figure*}
\includegraphics[width=0.99\textwidth,clip]{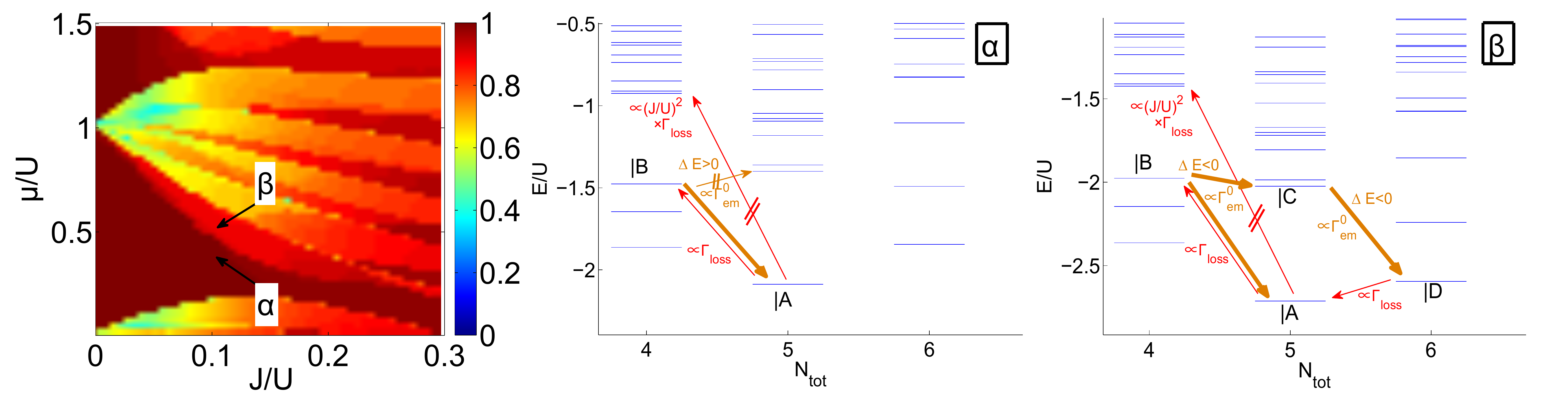}
\caption{\label{fig:spectrum_analysis} Left panel: Occupancy $\pi_{0}$ of the most populated quantum state at steady state for a 5 sites system. Central (resp. right) panel: spectrum of $H_{\text{eff}}$ at the values $J/U=0.1$ and $\mu/U=0.38$ (resp. $\mu/U=0.55$) indicated by the  point [$\alpha$] (resp. [$\beta$]) in the left panel. Parameters used: $\Gamma_{\rm{l}}/\Gamma_{\rm{em}}=10^{-3}$,  $U/\Delta_{\text{em}}=10^6$, $\Gamma_{\rm{em}}^0/\Delta_{\text{em}}=10^{-2}$ , $\omega_{+}=0$, $\omega_{-}/\Delta_{\text{em}}=-4\,.10^7$. }
\end{figure*}

This effect can be understood as a consequence of the departure of the pump and loss rates from a true detailed balance relation: the emission spectrum is in fact not exponential in the frequency but rather decays with a power law above $\omega_{+}$ and most importantly saturates at the value $\Gamma_{\rm{em}}^0$ below $\omega_{+}$. A physical interpretation of the underlying microscopic mechanism is illustrated in Fig.\ref{fig:spectrum_analysis}. In the left panel, we plot the occupancy $\pi_0$ of the most populated quantum state of the density matrix at steady state. In the central and right panels, we show the spectrum of the underlying Hamiltonian evaluated at two points [$\alpha$] and [$\beta$] separated by a small variation of $\mu$ for which the entropy is respectively zero and non-zero.

From the previous discussion, one expects that the steady state occupation be concentrated in the ground state $\ket{GS}=\ket{A}$ of $H_{\text{eff}}$, i.e., a (weakly delocalized) Mott state with 1 photon per site.  However looking at the spectrum of many-body quantum states for the choice of parameters indicated as $[\beta]$ (right panel of Fig.~\ref{fig:spectrum_analysis}),  we note that, starting from $\ket{A}$ which contains $N_{tot}=5$ photons in total, the system can lose one photon and arrive in a state $\ket{B}$ with $N_{tot}-1$ photons containing one hole excitation. Then, the pump re-injects a new photon and brings the system into a  doublon-hole excited quantum state  $\ket{C}\neq\ket{A}$ with $N_{tot}$ photons such that $E_{C}<E_{B}$. Since the spectrum has a square shape, the pump can bring the system toward both the ground state $\ket{A}$ and $\ket{C}$ with comparable efficiencies $\propto \Gamma_{\rm{em}}^0$.   There is one last doublon excited state $\ket{D}$ with  $N_{tot}+1$ photons and energy $E_{D}$ such that $E_{D}<E_{C}$ so the excited  $N_{tot}$ photon state  $\ket{C}$ is unstable and gets quickly pumped toward $\ket{D}$ where it gets trapped for a while as no state with higher photon number and lower energy exists, until one photon gets slowly lost and the system goes back to the ground state $\ket{A}$.  

This mechanism explains why we observe, in the steady state, a significantly non-zero entropy as well as a photonic density slightly bigger than in the ground state. Of course, if the emission rate were exponentially dependent in the energy jump and the detailed balance condition was verified, the re-pumping process toward $\ket{C}$ would not be relevant since its efficiency will be dynamically overwhelmed by the process bringing the system back toward the ground state $\ket{A}$. As a result, no significant trapping of population into excited states would occur.

In contrast, for the choice of parameters indicated as [$\alpha$] and illustrated in the central panel of Fig.~\ref{fig:spectrum_analysis}, the ground state $\ket{A}$ is well isolated dynamically. Looking at the spectrum of many-body quantum states shown in the right panel of Fig.~\ref{fig:spectrum_analysis}), one sees that the only energetically authorized transition after losing one photon is to go back into the Mott ground state of $N_{tot}$ photons. Of course there exist states with $N_{tot}-1$ photons in an higher energy band which would allow the kind of processes described earlier. However those states correspond to highly excited states (e.g., an hole combined with a doublon-hole) and have a much smaller overlap $\propto J/U$ with the state $a_{i}\ket{A}$ in which we removed one photon to the ground state. The effective rate of this process is thus of the order of $(J/U)^2 \Gamma_{\rm{l}}\sim \Gamma_{\rm{l}}/100$ and induces a negligible leak out of $\ket{A}$. As a consequence the steady state is almost pure, and corresponds very well to the Mott-like ground state. The sharpness of the transition between the two regimes at the $\alpha,\beta$ points is set by the edge linewidth of the order of $\Delta_{\rm em}=\Gamma_{\rm{p}}\ll U $.

Even though our interpretation of the effect is related to the level crossing between discrete hole and doublon-hole excited states of a finite system, we expect that a similar effect will occur in the thermodynamic limit  when the continuous energy bands of the hole excitations and doublon-hole excitations start to overlap above some tunneling $J=J_c$. Such an overlap can of course not happen when $J$ is relatively small with respect to $U$  since, for the first lobe for example, the hole band is separated from the ground-state by an energy $E^{hole}\simeq\mu\leq U$ while the doublon-hole one is separated by $E^{hole}_{doublon}\simeq U$. As a consequence, a critical value $J_c$ exist for this non-equilibrium channel to open up. Below this value, the Mott phase is expected to remain robust even in the thermodynamic limit. 

Whether this unexpected feature will affect the phase diagram in a dramatic manner (e.g., by destabilizing ordered phases~\cite{Altman15} and/or giving rise to novel exotic ones~\cite{Jin_davide1,Jin_davide2,Jin_alberto}) is a complex question that goes beyond the scope of this work and will be addressed in forthcoming works. Hints towards such exciting new physics are found in analytical calculations for generic non-equilibrium models~\cite{Altman15} and in the extensive numerics for the case of a Lorentzian emission spectrum in~\cite{Alberto}.

\section{An improved scheme for a full quantum simulation of the ground-state}
\label{sec:frequency-dependent_losses}
In this final section we introduce a further extension of the non-Markovian model of Sec.~\ref{sec:model} with the specific purpose of countering the effect of the non-equilibrium processes presented in Sec.~\ref{sec:non-equilibrium_features}, which induce a probability leakage out of the ground-state. This improved scheme is based on the introduction of extra frequency-dependent losses in addition to the natural Markovian ones.

\begin{figure}
\includegraphics[width=1\columnwidth]{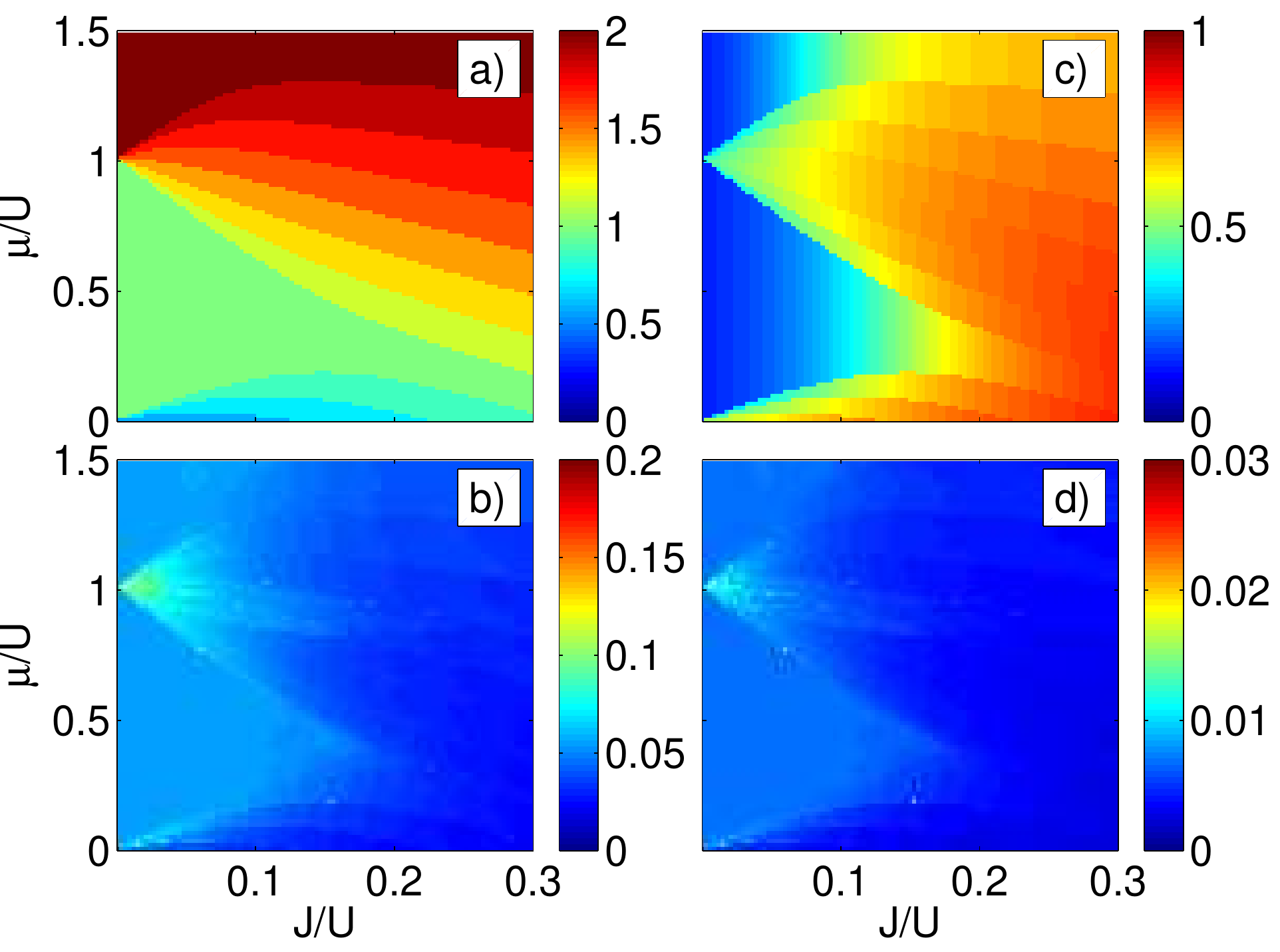}
\caption{\label{fig:frequency_dependent_losses}Steady-state properties using frequency-dependent losses with a square-spectrum in addition to frequency-dependent emission. Simulations were done for a $L=7$ sites periodic chain. Panel a) (resp. panel c): Average steady-state photon number per site $n_{\rm{ph}}$ (resp. condensed fraction $x_{\rm{BEC}}$). Panel b): steady-state entropy. Panel d): $1-\mathcal{F}$, where $\mathcal{F}=\bra{GS}\rho_\infty\ket{GS}$ is the fidelity of the steady-state density matrix $\rho_{\infty}$ with the ground-state $\ket{GS}$ of the Hamiltonian $H_{\rm{eff}}$, i.e, a $T=0$ state of chemical potential $\mu$. Same parameters as in Fig.~\ref{fig:phase_diagram} except for the additional frequency-dependent losses: $\Gamma^{0}_{\rm{L}}=\Gamma^{0}_{\rm{em}}$, $\omega_{\rm{L}}-\omega_{\rm{cav}}=\omega_{+}-\omega_{-}$ and $\Delta_{\rm{L}}=\Delta_{\rm{em}}$.}
\end{figure}
\subsection{The model}
We consider the following dynamics for the photonic density matrix
\begin{equation}
\partial_{t}\rho(t) =  -i\left[H_{\rm ph},\rho(t)\right]+\mathcal{L}_{\rm{l}} \big[\rho(t)\big]+ \mathcal{L}_{\rm{em}}\big[\rho(t)\big]+\mathcal{L}^{(\text{add})}_{\rm{L}} \big[\rho(t)\big],
\label{eq:photon_only2}
\end{equation}
where the Hamiltonian and dissipative contributions $H_{\rm ph}$, $\mathcal{L}_{\rm{l}} \big[\rho(t)\big]$ and $\mathcal{L}_{\rm{em}}\big[\rho(t)\big]$ are left unchanged with respect to Sec.~\ref{sec:model}. 

Similarly to emission, the additional frequency-dependent loss term
\begin{equation}
\label{eq:loss-non-markov}
\mathcal{L}^{(\text{add})}_{\rm{L}} \big[\rho(t)\big] = \frac{\Gamma_{\rm{L}}^{0}}{2}\sum_{i=1}^{L}\left[\bar{a}_{i}\rho a_{i}^{\dagger}+a_{i}\rho\bar{a}_{i}^{\dagger}-a_{i}^{\dagger}\bar{a}_{i}\rho-\rho\bar{a}_{i}^{\dagger}a_{i}\right]\!.
\end{equation}
involves modified lowering ($\bar a_i$) and raising ($\bar a^\dagger_i\equiv [\bar a_i]^\dagger $) \\operators
\begin{equation}
\label{eq:special-operators2}
\frac{\Gamma^{0}_{\rm{L}}}{2}\bar{a}_{i} = \!\int_{0}^{\infty}d\tau\, \Gamma_{\rm{L}}(\tau)a_{i}(-\tau),
\end{equation}
where 
\begin{equation}
\label{eq:memory-kernel2}
\Gamma_{\rm{L}}(\tau)=\theta(\tau)\int \frac{d\omega}{2\pi}\mathcal{S}_{\rm{L}}(\omega)e^{-i\omega\tau}.
\end{equation}
$\mathcal{S}_{\rm{L}}= s_{\rm L}(\omega) \Gamma^{0}_{\rm L}$ is the frequency-dependent loss rate, which  we also choose to be of a square shape as the emission term of
Sec.~\ref{sec:model}, by setting
\begin{equation}
s_{\rm{L}}(\omega)=\mathcal{N}'\,\int_{\omega_{+}}^{\omega_{\rm{L}}} d\omega' \frac{\Delta_{\rm{L}}/2}{(\omega-\omega')^2+(\Delta_{\rm{L}}/2)^2}
\label{eq:square_spectrum2}
\end{equation}
where the normalization constant $\mathcal{N}'$ is set such that $s_{\rm{L}}\left(\frac{\omega_++\omega_{\rm{L}}}{2}\right)=1$. Note the different choice for the loss frequency domain $[\omega_{+}, \omega_{\rm{L}}]$ (instead of $[\omega_{-}, \omega_{+}]$ for emission).

Such a kind of frequency-dependent losses can be implemented in a analogous manner to what was proposed in Sec.~\ref{sec:exp_proposal}: instead of using inverted emitters with a strong pumping toward the excited state, a possibility would be to couple our system to absorbers (or photonic resonators) with transition frequencies uniformly distributed over $[\omega_{+}, \omega_{\rm{L}}]$, and a very strong dissipative decay $\Gamma_{\downarrow}=\Delta_{\rm{L}}$ toward the ground-state (resp. vacuum state). Alternatively, one could couple the system to a single absorber (resp. resonator), whose transition frequency is temporally modulated over the interval $[\omega_{+}, \omega_{\rm{L}}]$.

We will choose strongly enhanced frequency-dependent losses with respect to the Markovian ones: $\Gamma^{0}_{\rm{L}}\gg\Gamma_{\rm{l}}$. Similarly to the emission spectrum, the upper cutoff $\omega_{\rm{L}}$ is not the most important feature and will be set to a very  far blue-detuned frequency: $\omega_{\rm{L}}-\omega_{\rm{cav}}\gg U,J\geq 0$. Likewise, we will have $\Delta_{\rm{L}}\ll U,\,\omega_{\rm{L}}-\omega_{+}$. All these conditions can be naturally satisfied, e.g., by mimicking the choice of parameters for emission: $\Gamma^{0}_{\rm{L}}=\Gamma^{0}_{\rm{em}}$, $\omega_{\rm{L}}-\omega_{+}=\omega_{+}-\omega_{-}$ and $\Delta_{\rm{L}}=\Delta_{\rm{em}}$. 

\subsection{Steady-state properties}
\label{subsec:steady-state}

In analogy to the frequency-dependent emission, the main effect of the frequency-dependent losses is to strongly enhance transitions removing a photon with a frequency above $\omega_{+}$. As a result, both non-Markovian emission and loss processes strongly accelerate transitions between many-body eigenstates which reduce the total energy computed using the effective Hamiltonian of Eq.~(\ref{eq:effective_hamiltonian}). Thus, the only quantum state for which both emission and losses are strongly suppressed (i.e., for which only natural Markovian losses are present) is the ground-state $\ket{GS}$ of $H_{\rm{eff}}$ (with $N_{\rm{tot}}$ photons), since it does not have states with $N_{\rm{tot}}-1$ and $N_{\rm{tot}}+1$ photons with lower energy.  

As a consequence, the ground-state $\ket{GS}$ has a long life time $\sim 1/\Gamma_{\rm{l}}\gg 1/\Gamma_{(\rm{em}/\rm{L})}^0$, while all remaining eigenstates have a short life-time $\sim 1/\Gamma_{(\rm{em}/\rm{L})}^0$. This important property was not ensured by the original scheme introduced in Sec.~\ref{sec:model}, for which some lowest-excited states with $N_{\rm{tot}}+1$ photons (e.g. the $\ket{D}$ state of Fig.~\ref{fig:spectrum_analysis} ) were long lived and could only relax  with a slow rate $\sim\Gamma_{\rm{l}}\ll\Gamma_{\rm{em}}^0$ towards $\ket{GS}$. This is the main reason for which this scheme shows a significant entropy in some regions of the parameter space. The new scheme including frequency-depedent losses solves this issue and is expected to be well suited to efficiently stabilize the ground state independently of the system parameters.

This statement based on simple physical arguments is confirmed in Fig.~\ref{fig:frequency_dependent_losses} where we see  that the steady-state average values of $n_{\rm{ph}}$ and $x_{\rm{BEC}}$ [panels a) and c)] are completely undistinguishable from the $T=0$ predictions of Fig.~\ref{fig:phase_diagram} [panels g) and h)].  We checked that this was also the case for higher order correlations. Even more remarkably, the steady-state $\rho_{\infty}$ has  a very low entropy [panel b)], and its fidelity $\mathcal{F}$ with the ground-state is very close to unity [panel d)] for any choice of parameters $\mu$ and $J$, indicating thus that we are indeed stabilizing a pure quantum state coinciding with the ground-state: $\rho_{\infty}=\ket{GS}\bra{GS}$.  

In contrast with the original scheme, there appears to be no real physical limitations to how close the steady-state can be to the ground-state $\ket{GS}$. We have in fact verified that the very small non-vanishing values for entropy (between $0.05$ and $0.12$) and deviations of the fidelity $\mathcal{F}$ from unity  (between $0.002$ and $0.012$) were a mere consequence of the finite choice of the dissipative parameters, and could be further reduced by orders of magnitude by improving the frequency selectivity $\Delta_{(\rm{em}/\rm{L})}/U$ of emission and losses and the ratios $\Gamma_{(\rm{em}/\rm{L})}/\Gamma_{\rm{l}}$. 

The fact that this improved scheme succeeds to stabilize the ground-state of the Bose-Hubbard model everywhere in the $\{\mu/U,J/U\}$ parameter space independently of the details of the underlying many-body physics (which is significantly different in the $J\ll U$ or $J\gg U$ cases) is a strong indication of its robustness and flexibility. We are therefore confident that this scheme can be efficiently applied to the quantum simulation of the zero temperature physics of a wide range of Hamiltonians. 

\section{Conclusions} 
\label{sec:conclusion}
In this work, we have introduced a novel pump scheme that allows to generate a gas of strongly interacting photons in driven-dissipative photonic systems and cool it down towards incompressible states. In particular, starting from a dissipative Bose-Hubbard model and using state-of-the-art parameters in circuit QED, we have demonstrated the feasibility of stabilizing Mott-Insulator-like states which are robust against tunneling and losses, and can be reshaped into coherent superfluid-like states for a suitable variation of parameters.

Depending on the specific values of the system and pumping parameters, the system behaviour can either closely resemble its equilibrium counterpart or show an unexpected transition to a non-equilibrium state characterized by a significant entropy. A strategy to circumvent this feature by adding frequency-dependent losses is proposed, and its efficiency to recover the Bose-Hubbard ground state as a steady state characterized. From an experimental perspective, simplification strategies to implement our proposal with just a few emitters/absorbers in a large lattice are pointed out.

In addition to observing the superfluid-insulator transition in a fluid of strongly interacting photons, our work demonstrates the possibility of quantum simulating zero-temperature equilibrium physics on a photonic platform. Future work will explore the possibility of exploiting the very non-equilibrium features to generate exotic many-body states and novel non-equilibrium phase transitions, as well as investigate the potential of our non-Markovian schemes with frequency-dependent pumping and losses to quantum simulate a wider range of many-body problems.

\acknowledgments Discussions with Atac Imamoglu, Jonathan Simon, Jamir Marino and Leonardo Mazza are warmly acknowledged.  JL and IC are supported by the EU-FET Proactive grant AQuS, Project
No. 640800, and by the Autonomous Province of Trento, partially through the project ``On silicon chip quantum optics for quantum computing and secure communications" (``SiQuro").  AB, FS and CC acknowledge support from ERC (via Consolidator Grant CORPHO No. 616233). RF acknowledges support from EU-QUIC. AB, CC, DR, and IC acknowledge the Kavli Institute for Theoretical Physics, University of California, Santa Barbara (USA) for the hospitality and support during the early stage of this work.

\appendix 
\section{Analytical expression for the emission spectrum shape}
\label{app:analytical_spectrum}
Here we give the precise analytical expressions for the emission spectrum $\mathcal{S}_{\rm{em}}(\omega)=\Gamma_{\rm{em}}^0 s_{\rm{em}}(\omega)$. It is defined as the convolution product between the square-shape function $s_{square}(\omega)=\theta(\omega-\omega_-)\theta(\omega_+-\omega)$ and a Lorentzian function of width $\Delta_{\rm{em}}$ : $s_{\rm{em}}=A(\omega)/A\left(\frac{\omega_{+}+\omega_{-}}{2}\right)$ where
\begin{eqnarray}
A(\omega)&=&\int_{\omega_{-}}^{\omega_{+}} d\omega' \frac{\Delta_{\rm{em}}/2}{(\omega-\omega')^2+(\Delta_{\rm{em}}/2)^2}\\
&=&\left[\text{arctan}\left(\frac{\omega_{+}-\omega}{\Delta_{\rm{em}}/2}\right)-\text{arctan}\left(\frac{\omega_{-}-\omega}{\Delta_{\rm{em}}/2}\right)\right].\nonumber
\end{eqnarray}
With this expression it is possible to compute the emission memory kernel 
\begin{equation}
\Gamma_{\rm{em}}(\tau)=\frac{i\Gamma_{\rm{em}}^0}{4}\frac{\theta(\tau)}{\tau}\frac{e^{(-i\omega_{+}-\Delta_{\rm{em}}/2)\tau}-e^{(-i\omega_{-}-\Delta_{\text{em}}/2)\tau}}{\text{arctan}\left(\frac{\omega_{+}-\omega_{-}}{\Delta_{\rm{em}}}\right)},
\end{equation}
as well as its Fourier transform
\begin{equation}
\Gamma_{\rm{em}}(\omega)= \frac{\Gamma_{\rm{em}}^0 }{2}s_{\rm{em}}(\omega)-i\delta_{\rm{lamb}}(\omega),
\end{equation}
where the frequency-dependent lamb-shift is given by
\begin{equation}
\delta_{\rm{lamb}}(\omega)=\frac{\Gamma_{\rm{em}}^0 }{2}\frac{\text{log}\left(\frac{(\omega_+-\omega)^2+(\Delta_{\rm{em}}/2)^2}{(\omega_--\omega)^2+(\Delta_{\rm{em}}/2)^2}\right)}{4\,\text{arctan}\left(\frac{\omega_{+}-\omega_{-}}{\Delta_{\rm{em}}}\right)}.
\end{equation}
When a transition frequency come close to the upper edge $\omega\simeq\omega_+$, the logarithm contribution can lead to an increase of the Lamb shift with respect to the power spectrum $S_{\rm{em}}(\omega)\sim \Gamma_{\rm{em}}^0$. However the logarithm diverges very slowly close to its singularities, and for our range of parameters (in the weakly dissipative regime) we have that the Lamb shift is at most $\delta_{\rm{lamb}}^{max}\sim 10\Gamma_{\rm{em}}^0\ll\Delta_{\rm{em}},U...$ and can thus be safely neglected.

\section{Derivation of the projected photonic master equation starting from a microscopic model}
\label{app:projective}
In this section, we give more details on the derivation of the photonic master equation introduced in the beginning of this letter, starting from the microscopic model proposed to engineer the pump. We focus for simplicity on the case of one cavity with a single embedded two-level emitter. Starting from the full emitter-cavity master equation, we show how  for a sufficiently small emitter-cavity coupling $\Omega_R$ the emitter degrees of freedom can be eliminated. The frequency-dependence of the amplification is then accounted for as a modified Lindblad term. Our treatment is based on the discussion in the textbook~\cite{Breuer}.

\subsection{\emph{General formalism}}
\label{app:proj_gen}
We consider a quantum system which undergoes dissipative processes. As it is not isolated, its state can not be described by a wave function but by a density matrix $\rho$ evolving according to the master equation:
\begin{equation}
\partial_{t}\rho=\mathcal{L}(\rho(t)),
\end{equation}
where $\mathcal{L}$ is some linear ``super-operator'' acting on the space of density matrices. Given an arbitrary initial density matrix $\rho(t_0)$, the density matrix $\rho$ at generic time $t$ 
is equal to $\rho(t)=e^{\mathcal{L}(t-t_{0})}\rho(t_{0})$.\\

Now we are only interested in some part of the density matrix, which
can represent some subsystem. This can be described by a projection
operation on the density matrix $\mathcal{P}\rho$ . We call $\mathcal{Q}=1-\mathcal{P}$
the complementary projector. We decompose the Lindblad operator $\mathcal{L}$ in two parts $\mathcal{L}_{0}$ and $\delta\mathcal{L}$ such that:
\begin{equation}
\left\{ \begin{array}{l}
\mathcal{\mathcal{L}=\mathcal{L}}_{0}+\delta\mathcal{L}\\
\mathcal{P}\mathcal{L}_{0}\mathcal{Q}=\mathcal{Q}\mathcal{L}_{0}\mathcal{P}=0\\
\mathcal{P}\,\delta\mathcal{L}\,\mathcal{P}=0.
\end{array}\right.\label{eq:condition projector}
\end{equation}
Such a decomposition is always possible. 

Then we define a generalised interaction picture for the density matrix and for generic superoperators $\mathcal{A}$ with respect to the evolution described by the free $\mathcal{L}_{0}$ and the initial time $t_{0}$: 
\begin{equation}
\left\{ \begin{array}{l}
\hat{\rho}(t)=e^{-\mathcal{L}_{0}(t-t_{0})}\rho(t)\\
\hat{\mathcal{\mathcal{A}}}(t)=e^{-\mathcal{L}_{0}(t-t_{0})}\mathcal{A}e^{\mathcal{L}_{0}(t-t_{0})}.
\end{array}\right.
\end{equation}

As discussed in \cite{Breuer}, we can get an exact closed master
equation for the projected density matrix in the interaction picture
\begin{equation}
\partial_{t}\mathcal{P}\hat{\rho}(t)=\int_{t_{0}}^{t}dt'\Sigma(t,t')\mathcal{P}\hat{\rho}(t'), \label{eq: ev self int}
\end{equation}
which translates to 
\begin{eqnarray}
\partial_{t}\mathcal{P}\rho(t) & = & \mathcal{L}_{0}(\rho(t))+\int_{t_{0}}^{t}dt'\tilde{\Sigma}(t-t')\mathcal{P}\rho(t')\label{eq:ev self schro-1}
\end{eqnarray}
in the Schrodinger picture. In the interaction picture, the self energy operator $\Sigma$ is defined as:
\begin{eqnarray}
\Sigma(t,t') &=& \sum_{n=2}^{\infty}\int_{t'}^{t}\int_{t'}^{t_{1}}..\int_{t'}^{t_{n-1}}dt_{1}..dt_{n}\\
&&\mathcal{P}\delta\hat{\mathcal{L}}(t)\mathcal{Q}\delta\hat{\mathcal{L}}(t_{1})\mathcal{Q}\delta\hat{\mathcal{L}}(t_{2})...\mathcal{Q}\delta\hat{\mathcal{L}}(t_{n})\mathcal{Q}\delta\hat{\mathcal{L}}(t')\mathcal{P}\nonumber
\end{eqnarray}
and results from the coherent sum over the processes leaving from $\mathcal{P}$,
remaining in $\mathcal{Q}$ and then coming back finally to $\mathcal{P}$.
In the Schrodinger representation, we have:
\begin{equation}
\tilde{\Sigma}(t-t')=e^{\mathcal{L}_{0}(t-t_{0})}\Sigma(t,t')e^{-\mathcal{L}_{0}(t'-t_{0})}=\Sigma(0,t'-t)e^{\mathcal{L}_{0}(t-t')}.\label{app:schrodinger-self-energy}
\end{equation}

We call $\tau_{c}=1/\Delta\omega$ the characteristic decay time /
inverse linewidth for the self energy, which corresponds in general to
the correlation time of the bath, and we estimate the rate of dissipative processes as $\Gamma\sim \text{max}_{\omega} |\tilde{\Sigma}(\omega)|$. We
put ourselves in the regimes in which, with respect to these dissipative
processes, the bath has a short memory, ie $\Gamma\ll\Delta\omega$.
In that regime the density matrix in the interaction picture is almost
constant over that time $\tau_{c}$. Furthemore, if $t-t_{0}\gg\tau_{c}$
then the integral in eq (\ref{eq: ev self int}) can be extended from $-\infty$
to $t$. From this equation and from (\ref{eq: ev self int}), by going back in the Schrodinger picture we get an
equation of evolution for the density matrix which is local in time:
\begin{equation}
\partial_{t}\mathcal{P}\hat{\rho}(t) = \left[\mathcal{L}_{0}+\int_{0}^{\infty}d\tau\Sigma(0,-\tau)\right]\,\mathcal{P}\rho(t)=
\mathcal{L}_{\text{eff}}\mathcal{P}\rho(t),\label{eq:approx eff ev}
\end{equation}
with
\begin{equation}
\mathcal{L}_{\text{eff}}=\mathcal{L}_{0}+\int_{0}^{\infty}d\tau\Sigma(0,-\tau).
\end{equation}
It is worth stressing that while the bath is Markovian with respect to dissipative processes induced by the perturbation $\int_{0}^{\infty}d\tau\Sigma(0,-\tau)$ , no Markovian approximation has been made with respect to the dynamics due to $\mathcal{L}_{0}$, which can still be fast. For the specific system under consideration in this work, this means that the emission rate $\Gamma_{\rm{em}}$ has to be slow with respect to the gain bandwidth set by the emitter pumping rate $\Gamma_{\rm{p}}$, which is the case in the weak coupling limit $\sqrt{N_{\rm{at}}}\Omega_{R}\ll\Gamma_{\rm{p}}$. However no restriction is to be imposed on the parameters $U$, $J$ and $\omega_{\rm{cav}}-\omega_{\rm{at}}$ of the Hamiltonian, which can be arbitrarily large. This means that the physics can be strongly non-Markovian with respect to the Hamiltonian photonic dynamics.

\subsection{\emph{Application to the array of cavities} }
\subsubsection*{Preliminary calculations}

With the notation of our \{emitters+cavity modes\} proposal of implementation, we choose
the projectors in the form:
\begin{equation}
\mathcal{P}\rho=\ket{\{e_{i}^{(n)}\}}\bra{\{e_{i}^{(n)}\}}\otimes \rm{Tr}_{\rm{at}}(\rho),
\end{equation}
where we have performed a partial trace over the embedded emitters in all cavities, and then make the tensor
product of the density matrix and the emitter density matrix with all
emitters in the excited state. We chose this particular projector because in the weak emitter-cavity coupling regime, we expect emitters to be repumped almost immediately after having emitted a photon in the cavity array, and thus to be most of the time in the excited state. Moreover this projection operation gives us direct access to the photonic density matrix, and thus we do not lose any information on photonic statistics. Applying the method sketched in the previous section, and restricting ourselves to the single emitter-single cavity configuration,  we derive the following photonic master equation (see \cite{Lebreuilly} for the details of the derivation):
\begin{eqnarray}
\partial_{t}\rho &=&  -i\left[H_{\rm{ph}},\rho_{ph}\right]+\frac{\Gamma_{\rm{l}}}{2}\left[2a\rho a^{\dagger}-a^{\dagger}a\rho-\rho a^{\dagger}a\right] \nonumber\\
&&+\frac{2\Omega_{R}^{2}}{\Gamma_{\rm{p}}}\left[\tilde{a}^{\dagger}\rho a+a^{\dagger}\rho\tilde{a}-a\tilde{a}^{\dagger}\rho-\rho\tilde{a}a^{\dagger}\right],
\end{eqnarray}
with 
\begin{equation}
\left\{ \begin{array}{l}
\tilde{a}=\frac{\Gamma_{\rm{p}}}{2}\int_{0}^{\infty}d\tau\, e^{(-i\omega_{\rm{at}}-\Gamma_{\rm{p}}/2)\tau}a(-\tau) ,
\\
\tilde{a}^{\dagger}=\frac{\Gamma_{\rm{p}}}{2}\int_{0}^{\infty}d\tau\, e^{(i\omega_{\rm{at}}-\Gamma_{\rm{p}}/2)\tau}a^{\dagger} (-\tau)=\left[\tilde{a}\right]^{\dagger} ,
\end{array}\right.
\end{equation}
where $ a(-\tau)$ means the photonic annihilation operator in the photonic Hamiltonian interaction picture. 

If $\ket f$ and $\ket f'$ are two eigenstates of the photonic Hamiltonian with a photon number difference of one, we see that the matrix elements of the modified annihilation and creation operators $\tilde{a}$ and $\tilde{a}^{\dagger}$ involved in the emission process are:
\begin{equation}
\left\{ \begin{array}{l}
\bra{f}\tilde{a}\ket {f'}=\frac{\Gamma_{\rm{p}}/2}{-i(\omega_{f'f}-\omega_{\rm{at}})+\Gamma_{\rm{p}}/2}\bra{f}a\ket {f'}\\
\bra{f'}\tilde{a}^{\dagger}\ket{f}=\frac{\Gamma_{\rm{p}}/2}{i(\omega_{f'f}-\omega_{\rm{at}})+\Gamma_{\rm{p}}/2}\bra{f'}a^{\dagger}\ket{f}.
\end{array}\right.
\end{equation}
The non-Markovianity comes from the energy-dependence of the prefactors. 

For this simple configuration and with the notations of the main text [Eq.~(4)], we have thus that $\Gamma_{\rm{em}}(\tau)=2\Omega_{R}^2 \theta(\tau)e^{(-i\omega_{\rm{at}}-\Gamma_{\rm{p}}/2)\tau}$ and $\Gamma_{\rm{em}}(\omega)=\frac{\Gamma_{\rm{em}}^{\rm{at}}}{2}\frac{\Gamma_{\rm{p}}/2}{-i(\omega-\omega_{\rm{at}})+\Gamma_{\rm{p}}/2}$, where $\Gamma_{\rm{em}}^{\rm{at}}=\frac{4\Omega_{R}^2}{\Gamma_{\rm{p}}}$ is the emission rate of a single emitter at the top of the Lorentzian.
\subsubsection*{Master equation for many cavities and many emitters}
For several cavities and a large number $N_{\rm{at}}$ of emitters per cavity (modeled by a continuum of bare-frequencies with the distribution $\mathcal{D}(\omega)$), the reasoning is exactly the same: each emitter brings its own inifinitesimal contribute to the total frequency-dependent emission, and by making the continuous sum of all of these terms we get the multicavity master equation
\begin{eqnarray}
\partial_{t}\rho &= & -i\left[H_{\rm{ph}},\rho(t)\right]+\mathcal{L}_{\rm{l}} (\rho(t))+\\
&& \frac{\Gamma_{\rm{em}}^0}{2}\sum_{i=1}^{L}\left[\tilde{a}_{i}^{\dagger}\rho a_{i}+a_{i}^{\dagger}\rho\tilde{a}_{i}-a_{i}\tilde{a}_{i}^{\dagger}\rho-\rho\tilde{a}_{i}a_{i}^{\dagger}\right],\nonumber
\label{eq:app-photon_only}
\end{eqnarray}
where
\begin{equation}
\frac{\Gamma^{0}_{\rm{em}}}{2}\tilde{a}_{i}=\int_{0}^{\infty}d\tau\, \Gamma_{\rm{em}}(\tau)a_{i}(-\tau) ,\qquad\qquad  \tilde{a}_i^{\dagger}=\left[\tilde{a}_i\right]^{\dagger}
\end{equation}
is the modified annihilation operator,
\begin{eqnarray}
\label{eq:app-emission-kernel}
\Gamma_{\rm{em}}(\tau)&=&\Gamma_{\rm{em}}^{\rm{at}}\theta(\tau)\int d\tilde{\omega}\mathcal{D}(\tilde{\omega})e^{-(i\tilde{\omega}+\Gamma_{\rm{p}}/2)\tau}\\
&=&\theta(\tau)\int \frac{d\omega}{2\pi}\mathcal{S}_{\rm{em}}(\omega)e^{-i\omega\tau}
\end{eqnarray}
is the causal autocorrelation for photonic emission,
\begin{equation}
\label{eq:app-power-spectrum}
\mathcal{S}_{\rm{em}}(\omega)=\Gamma_{\rm{em}}^{\rm{at}}\int d\tilde{\omega}\mathcal{D}(\tilde{\omega})\frac{(\Gamma_{\rm{p}}/2)^2}{(\omega-\tilde{\omega})^2+(\Gamma_{\rm{p}}/2)^2}
\end{equation}
is the photonic emission power spectrum and $\Gamma_{\rm{em}}^{\rm{at}}$ has been defined in the previous paragraph. $\mathcal{S}_{\rm{em}}(\omega)$ is the convolution product of a Lorentzian which represents the broadening of each emitter due to the pumping, and the spectral distribution $\mathcal{D}(\omega)$ of the emitter bare frequencies in absence of pumping. In our case, the distribution is square-shape  $\mathcal{D}^{square}(\omega)= \frac{N_{\rm{at}}}{\omega_{+}-\omega_{-}}\theta(\omega-\omega_{-})\theta(\omega_{+}-\omega)$, so we obtain the form for the emission power spectrum:
\begin{equation}
\mathcal{S}^{square}_{\rm{em}}(\omega)=\Gamma_{\rm{em}}^{\rm{at}}\frac{N_{\rm{at}}}{\omega_{+}-\omega_{-}}\int_{\omega_{-}}^{\omega_{+}}  d\tilde{\omega}\frac{(\Delta_{\rm{em}}/2)^2}{(\omega-\tilde{\omega})^2+(\Delta_{\rm{em}}/2)^2},
\end{equation}
with $\Delta_{\rm{em}}=\Gamma_{\rm{p}}$. The maximum power spectrum obtained at the middle between the two cutoffs is then
\begin{eqnarray}
\Gamma_{\rm{em}}^0&=&\mathcal{S}^{square}_{\rm{em}}\left(\frac{\omega_++\omega_-}{2}\right)\nonumber\\
&=&\frac{2\pi N_{\rm{at}}\Omega_{R}^2}{\omega_+-\omega_-}\qquad \rm{for} \,\,\,\Delta_{\rm{em}}\ll\omega_+-\omega_- 
\end{eqnarray}
 
\section{More on experimental scheme simplifications}
\label{app:scheme_simplifications}
\subsection{Pumping only a few sites}
\label{app:pumping_a_few_sites}
\begin{figure}
\includegraphics[width=1\columnwidth]{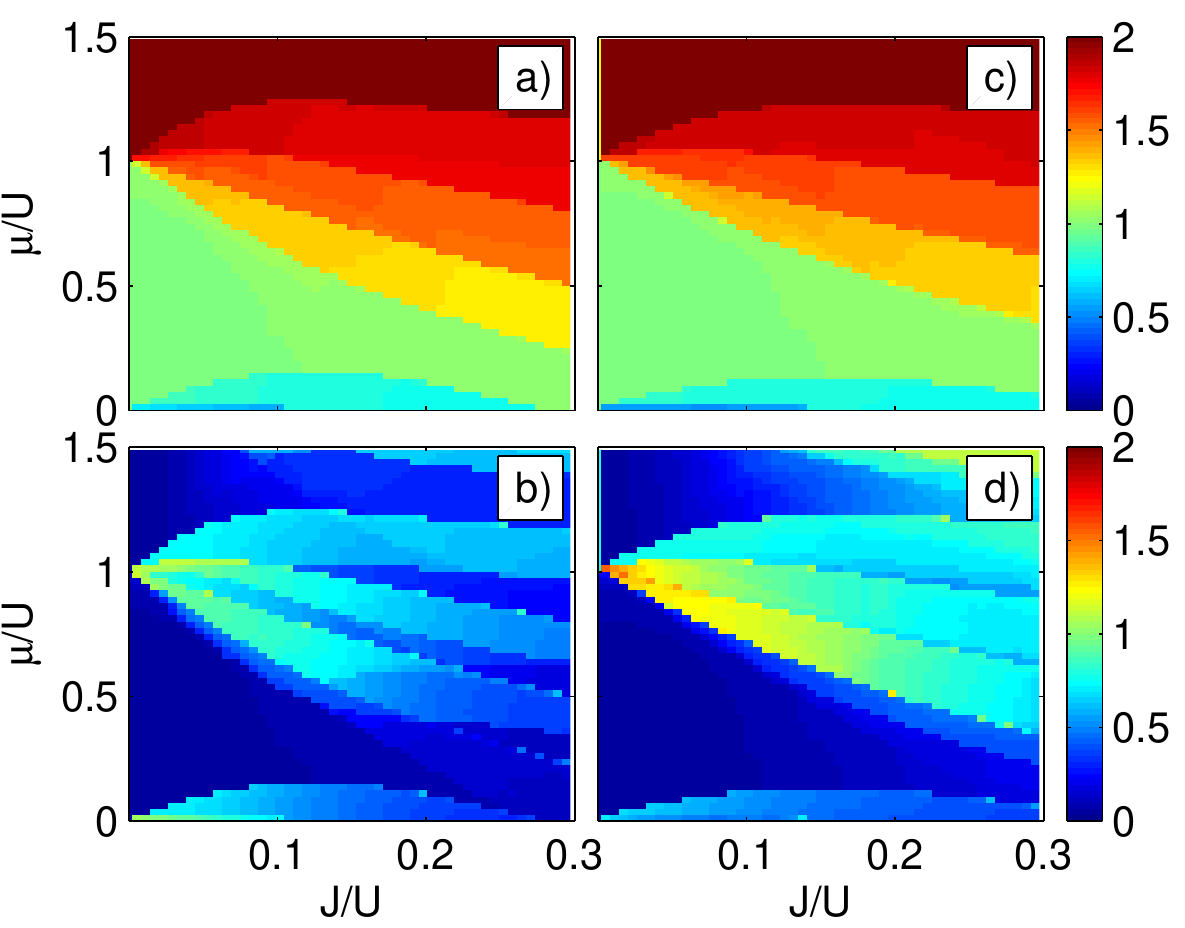}
\caption{\label{fig:single_double_emitter}Steady-state properties using only one or two emitting sites in different geometries. Panel a) (resp. c)) show the steady-state average density $n_{\rm{ph}}$, and panel b) (resp. d)) the entropy for a $L=4$ sites system with periodic boundary conditions with emitters localized on the first two sites (resp. an open chain with emitters only on first site). Same parameters as in Fig.~\ref{fig:phase_diagram} except for the stronger emission rate to compensate the reduced number of emitting sites: $\Gamma_{\rm{em}}^0/\Delta_{\rm{em}}=\frac{L}{2}\times 10^{-2}$ in panels a), b) (resp.  $L\times10^{-2}$ in panels c),d)). A smaller lattice of $L=4$ sites had to be used because of the broken translational invariance.}
\end{figure}

In this first part of this Appendix, we discuss further on the possibility of using only a few emitting sites in order to stabilize the desired steady-state of Sec.~\ref{sec:numerical_results}.  In order to justify this, let us understand what happens when a hole excitation is created by losing a photon starting from a state $\ket{f'}$ and arriving in a state $\ket{f}$. We want to understand how many emitting sites are needed to refill the many-body state back to the initial state $\ket{f'}$.  The rates of such processes are $\mathcal{T}_{f\to f'}=|\bra{f'}a_{i}\ket{f}|^2 \mathcal{S}_{em}(\omega_{f',f})$ where $i$ is the position of the emitting site, $\omega_{f',f}=\omega_{f'}-\omega_{f}$ is the energy cost needed to remove the hole excitation and the matrix element $\bra{f'}a_{i}\ket{f}$ represents the wave-function amplitude for finding the hole excitation on $i$. 

For very weak tunneling $J/U\ll1$, hole excitations behave like free particles. In a periodic chain, if the single-hole wave function had a complex plane-wave form, then the modulus of its amplitude would be fully position-independent and all emitting sites would lead to the the same refilling rate. However due to reflection symmetry, loss generated hole excitations possess a symmetric distribution of momenta and their wave packet has a cosine-shaped standing wave profile. As a consequence the amplitude of each of these waves is spatially modulated and presents nodes at fixed location in the lattice. If an emitting site is located on the node of such hole excitation, it can not feel at all the presence of the excitation, making the refilling impossible as $\bra{f'}a_{i}\ket{f}=0$.

\begin{figure}
\includegraphics[width=1\columnwidth]{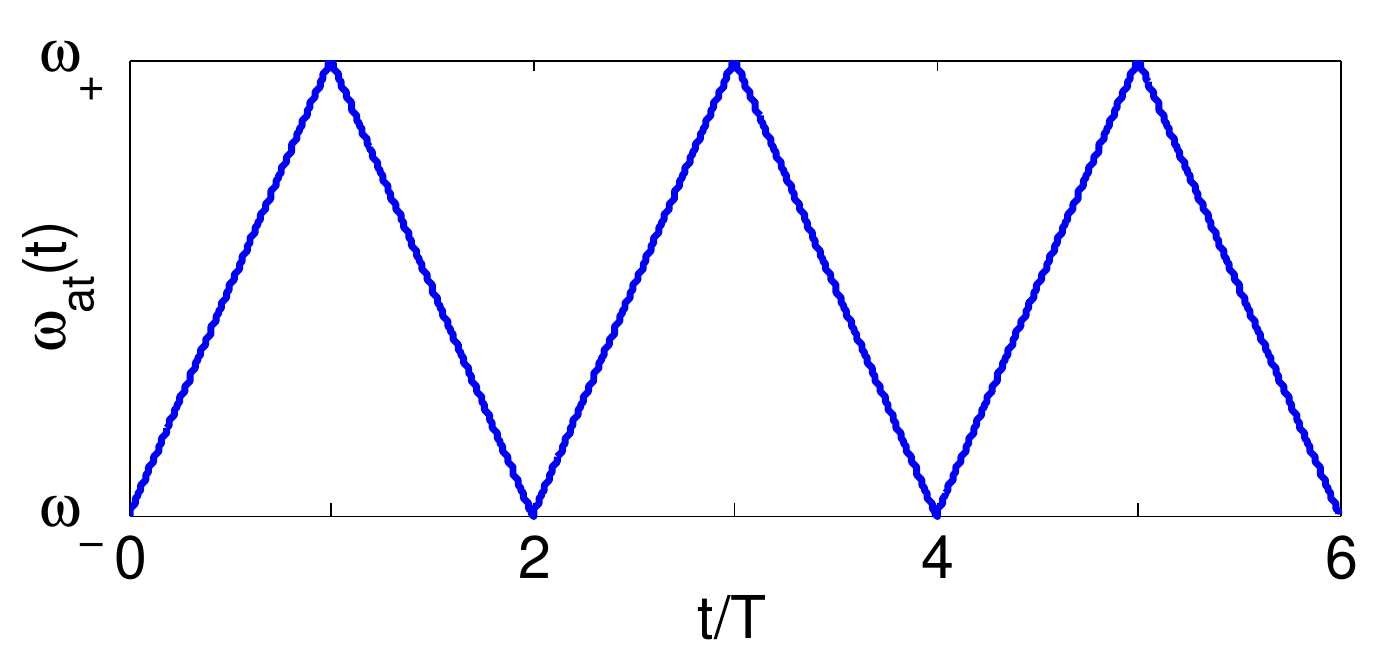}\vspace{-2mm}
\caption{\label{fig:frequency_modulation} Temporal profile of the time-dependent emitter transition frequency required to mimic the square-shaped emission spectrum}
\end{figure}

In a periodic chain, for only one emitting site located in $i$, one therefore expects that the steady-state will always maintain a finite number of hole excitations whose nodes are pinned at the emitting site, leading to a spatially modulated density profile with a maximum on the emitting site. However since a single-hole excitation standing-wave amplitude can never have two nodes on consecutive sites, using two nearest neighbors as emitting sites would prevent this effect, as either one or the other of the emitting sites would feel the presence of the hole and would replenish the many-body state, as confirmed in Fig.~\ref{fig:single_double_emitter} (panels a),b)).

One concludes that for a periodic chain (under the requirement  $J\gg \Gamma_{\rm{l}},\,\Gamma_{\rm{em}}^0$ which allows holes excitations to travel fast enough) only two nearest neighbor emitting sites are required to stabilize a Mott state of perfectly integer density and weak entropy, and thus to replace pumping on all sites, which is an important experimental simplification. 

\begin{figure*}
\includegraphics[width=1\textwidth]{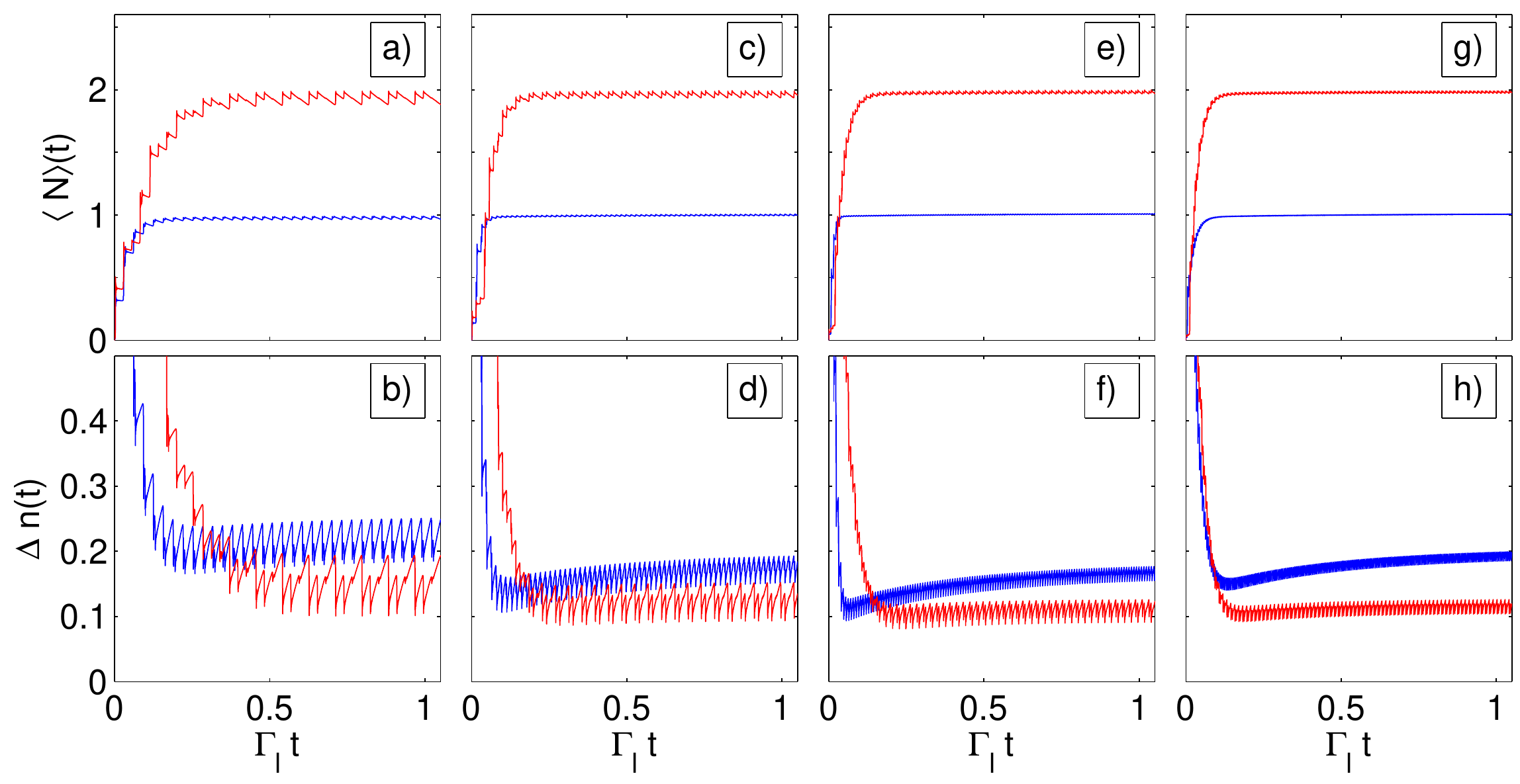}\vspace{-2mm}
\caption{\label{fig:dynamic_modulation} Time evolution of a single cavity configuration using a single temporally modulated emitter to mimic the effect of a square emission spectrum. Different panels from left to right refer to increasing values of the modulation speed $v_\omega=\left|\frac{d\omega_{at}}{dt}\right|$. The blue (resp. red) lines correspond to a single photon (resp. two photon) Fock state stabilization. Upper panels: average photon number $\left\langle N\right\rangle (t)$. Lower panels:  relative fluctuations of the total particle number  $\Delta n (t)=\sqrt{\left\langle N^2\right\rangle(t)-\left\langle N\right\rangle^2(t)}/\left\langle N\right\rangle(t)$. The photon-emitter Rabi coupling  $\Omega_{R}$ was chosen for each modulation speed in such a way to set the stationary value of the photon number close the desired occupation number: $\left\langle N\right\rangle(t)\simeq_{t\to\infty} 1$ (resp. $2$). Parameters inspired from state-of-the-art in circuit-QED systems~\cite{Ma_Simon,Rigetti}: $U= 200 \times2\pi\text{MHz}$, $\Gamma_{\rm{p}}=0.5 \times 2\pi\text{MHz}$, $\Gamma_{\rm{l}}=1\times 2\pi\text{kHz}$. For the blue (resp. red) lines $\mu=\omega_{+}-\omega_{\rm{cav}}=U/2$ (resp. $3U/2)$, and $\omega_{+}-\omega_{-}=0.6\,U$ (resp. $1.6\,U$). Choice for the modulation speed, from left to right: $v_{\omega}=7.5\times(2\pi\text{MHz})^2$, $15\times(2\pi\text{MHz})^2$, $30\times(2\pi\text{MHz})^2$, $50\times(2\pi\text{MHz})^2$. Corrispondingly for the blue lines, from left to right $\Omega_{R}=0.83\times 2\pi\text{Hz}$, $0.9\times 2\pi\text{Hz}$, $0.97\times 2\pi\text{Hz}$, $1.07\times 2\pi\text{Hz}$ (resp. for the red lines $\Omega_{R}=1\times 2\pi\text{Hz}$, $1.07\times 2\pi\text{Hz}$, $1.15\times 2\pi\text{Hz}$, $1.24\times 2\pi\text{Hz}$).}
\end{figure*}

In the open-chain configuration things are even simpler, as we know that a standing-wave single-hole excitation on a lattice never presents exactly a node at the first or last lattice site. As a consequence using only one site as an emitter at one of the extremities of the chain is sufficient [Fig.~\ref{fig:single_double_emitter} c),d)].

\subsection{Temporally modulated emitters}
\label{app:temporally_modulated}

In this second part of this Appendix, we provide some preliminary checks of the validity of the approach consisting in modulating the frequency of a single emitter/a few emitters (according to the time profile of Fig.~\ref{fig:frequency_modulation}) in order to mimic the effect of a square spectrum. In particular, we focus on the effect of the temporal modulation on the fluctuations in the steady-state. As a first step, in Fig.~\ref{fig:dynamic_modulation}  we proceeded to a simple test in a single-cavity configuration, investigating the possibility of stabilizing a single-photon (resp. two-photon) Fock state by setting $\mu=U/2$ (resp. $\mu=3U/2$) in the middle of the first (resp.second) Mott lobe, and we compared the resulting performance between several modulation speeds $v_\omega=\left|\frac{d\omega_{at}}{dt}\right|$. 
 
 Our results are consistent with the qualitative discussions of Sec.~\ref{sec:temporal_modulation}: for low modulation speeds (panels a)-d)), losses occuring within the modulation half-period $T=(\omega_+-\omega_-)/v_\omega$ of the modulated emitter can not be neglected, and thus the density maintains measurable oscillatory behaviour, a true steady-state is not fully reached and fluctuations are substantial. For an optimal modulation speed (panels e),f)) we obtained , using state-of-the art parameters of circuit QED, a minimized value  for the particle number relative fluctuations $\Delta n\simeq 0.17$ (resp. $0.12$) for the first (resp.second) Fock state, leading to a probability $\pi\simeq\left\langle N^2\right\rangle-\left\langle N\right\rangle^2$ of only $3\%$ (resp. $5\%$) of not being in the desired Fock state, and an effective temperature $T_{\rm{eff}}\simeq 0.12\times U$ (resp. $0.14\times U$): this is a very good level of performance, comparable to the results of Sec.~\ref{sec:sota} obtained by direct steady-state calculation of the master equation Eq.~(\ref{eq:photon_only}) (for which we obtained $\pi=1.7\%$ and $T_{\rm{eff}}\simeq 0.10\times U$ for the first lobe using similar parameters) . At higher modulation speeds  (panels g),h)), as discussed in Sec.~\ref{sec:temporal_modulation}, the modulation-induced broadening is responsible for an increase of fluctuations as it leads to heating effects and undesired transitions toward Fock states with higher photon number. 
 
As we can see, temporal fluctuations are slightly more important if we want to stabilize a Fock state with an higher photon number, since the emitter needs to travel over a broader range of frequencies, in order to protect all transitions from the vacuum until the desired occupation number, and thus losses are more important over the modulation time interval $T=(\omega_+-\omega_-)/v_\omega$. As was explained in \ref{sec:temporal_modulation}, this issue can be fixed by using several emitters spanning different frequency regions.

These preliminary tests on a very simplified single-cavity configuration confirm the promise of this method. A complete study including the complexity of a many-cavity array will be the subject of a future work.

\end{document}